\input harvmac 
\input epsf.tex

\overfullrule=0mm

\newcount\figno
\figno=0
\def\fig#1#2#3{
\par\begingroup\parindent=0pt\leftskip=1cm\rightskip=1cm\parindent=0pt
\baselineskip=11pt
\global\advance\figno by 1
\midinsert
\epsfxsize=#3
\centerline{\epsfbox{#2}}
\vskip 12pt
{\bf Fig. \the\figno:} #1\par
\endinsert\endgroup\par
}
\def\figlabel#1{\xdef#1{\the\figno}}
\def\encadremath#1{\vbox{\hrule\hbox{\vrule\kern8pt\vbox{\kern8pt
\hbox{$\displaystyle #1$}\kern8pt}
\kern8pt\vrule}\hrule}}

\def\wrt{with respect to\ }
\def\z{{\zeta}}

\def\IR{\relax{\rm I\kern-.18em R}}
\font\cmss=cmss10 \font\cmsss=cmss10 at 7pt

\font\cmss=cmss10 \font\cmsss=cmss10 at 7pt
\def\IZ{\relax\ifmmode\mathchoice
{\hbox{\cmss Z\kern-.4em Z}}{\hbox{\cmss Z\kern-.4em Z}}
{\lower.9pt\hbox{\cmsss Z\kern-.4em Z}}
{\lower1.2pt\hbox{\cmsss Z\kern-.4em Z}}\else{\cmss Z\kern-.4em Z}\fi}
\def\IN{\relax{\rm I\kern-.18em N}}


\Title{\vbox{\hsize=3.truecm \hbox{SPhT/02-093}}}
{{\vbox {
\bigskip
\centerline{Census of Planar Maps:}
\centerline{From the One-Matrix Model Solution}
\centerline{to a Combinatorial Proof}
}}}
\bigskip
\centerline{J. Bouttier\foot{bouttier@spht.saclay.cea.fr}, 
P. Di Francesco\foot{philippe@spht.saclay.cea.fr} and
E. Guitter\foot{guitter@spht.saclay.cea.fr}}
\medskip
\centerline{ \it Service de Physique Th\'eorique, CEA/DSM/SPhT}
\centerline{ \it Unit\'e de recherche associ\'ee au CNRS}
\centerline{ \it CEA/Saclay}
\centerline{ \it 91191 Gif sur Yvette Cedex, France}
\bigskip
\noindent 
We consider the problem of enumeration of planar maps
and revisit its one-matrix model solution in the 
light of recent combinatorial techniques involving
conjugated trees. We adapt and generalize these techniques
so as to give an alternative and purely combinatorial 
solution to the problem of counting
arbitrary planar maps with prescribed vertex degrees.

\Date{07/02}

\nref\TUTone{W. Tutte, 
{\it A Census of planar triangulations}
Canad. Jour. of Math. {\bf 14} (1962) 21-38.}
\nref\TUTtwo{W. Tutte, 
{\it A Census of Hamiltonian polygons}
Canad. Jour. of Math. {\bf 14} (1962) 402-417.}
\nref\TUTthree{W. Tutte, 
{\it A Census of slicings}
Canad. Jour. of Math. {\bf 14} (1962) 708-722.}
\nref\TUTfour{W. Tutte, 
{\it A Census of Planar Maps}, Canad. Jour. of Math. 
{\bf 15} (1963) 249-271.}
\nref\Tho{G. 't Hooft, {\it A planar diagram theory for strong
interactions}, Nucl. Phys. {\bf B72} (1974) 461-473.}
\nref\BIPZ{E. Br\'ezin, C. Itzykson, G. Parisi and J.-B. Zuber, {\it Planar
Diagrams}, Comm. Math. Phys. {\bf 59} (1978) 35-51.}
\nref\DGZ{P. Di Francesco, P. Ginsparg
and J. Zinn--Justin, {\it 2D Gravity and Random Matrices},
Physics Reports {\bf 254} (1995) 1-131.}
\nref\EY{B. Eynard, {\it Random Matrices}, Saclay Lecture Notes (2000),
available at {\sl http://www-spht.cea.fr/lectures\_notes.shtml} }
\nref\SCHrev{G. Schaeffer, {\it Conjugaison d'arbres
et cartes combinatoires al\'eatoires} PhD Thesis, Universit\'e 
Bordeaux I (1998).}
\nref\SCH{G. Schaeffer, {\it Bijective census and random 
generation of Eulerian planar maps}, Electronic
Journal of Combinatorics, vol. {\bf 4} (1997) R20.}
\nref\PoSc{D. Poulhalon and G. Schaeffer, {\it A bijection for loopless 
triangulations of a polygon with interior points}, proceedings of the 
conference FPSAC'02, Melbourne (2002),
available at {\sl http://www.loria.fr/$\sim$schaeffe/}}
\nref\BMS{M. Bousquet-M\'elou and G. Schaeffer,
{\it Enumeration of planar constellations}, Adv. in Applied Math.,
{\bf 24} (2000) 337-368.}
\nref\PS{D. Poulhalon and G. Schaeffer, 
{\it A note on bipartite Eulerian planar maps}, preprint (2002),
available at {\sl http://www.loria.fr/$\sim$schaeffe/}}
\nref\DEG{P. Di Francesco, B. Eynard and E. Guitter,
{\it Coloring Random Triangulations},
Nucl. Phys. {\bf B516 [FS]} (1998) 543-587.}
\nref\BFG{J. Bouttier, P. Di Francesco and E. Guitter,
{\it Counting colored Random Triangulations}, Saclay preprint SPhT/02-075,
cond-mat/0206452, to appear in Nucl. Phys. B.}
\nref\ISING{see for instance D. Boulatov and V. Kazakov, {\it The Ising model
on a random planar lattice: the structure of the phase 
transition and the exact critical exponents}, Phys. Lett. {\bf B186} (1987) 379-384.}
\nref\HARD{J. Bouttier, P. Di Francesco and E. Guitter, {\it Critical
and tricritical hard objects on bicolourable random lattices: exact solutions}
J. Phys. A: Math. Gen. {\bf 35} (2002) 3821-3854.}

\newsec{Introduction}

Enumeration of planar maps has been a classical subject
of combinatorics originally motivated by the famous four-color problem. 
Major advances in this field were obtained in the 60's 
by W. Tutte in his famous ``Census" papers [\xref\TUTone-\xref\TUTfour], 
giving many explicit enumerations for various classes of planar maps.
Fifteen years later, the same problem became popular
among physicists in the context of the perturbative 
expansion of SU($N$) gauge field theory \Tho. Indeed, at large $N$,
the dominant Feynman diagrams correspond precisely to planar maps. 
Explicit enumeration formulas extending previous results 
were derived by matrix integral techniques \BIPZ, a tool which proved 
very powerful for such problems. Even more recently, planar maps 
were used in physics as tessellations of random surfaces, in the
context of both discretized 2D quantum gravity and fluid membranes
(see e.g. \DGZ\ and \EY).

Planar maps are formally defined as proper embeddings of graphs
into the two-dimensional Riemann sphere, considered up to continuous
deformations. A map is characterized by a number of {\it vertices},
{\it edges} and {\it faces}. The general question we address here
is the enumeration all such maps with
prescribed vertex degrees, i.e, for each $k$, a fixed number $n_k$ 
of $k$-valent vertices. Turning to generating functions, we can 
instead consider the partition function of all maps with a weight 
$g_k$ per $k$-valent vertex. It proves also useful to introduce 
an additional weight $t$ per edge so that a term of given degree 
in $t$ accounts for a finite number of maps. 

The effective counting can be performed in a mechanical way
by solving the generic one-matrix model with standard techniques. 
This approach provides directly the net result, but its powerful 
nature may seem magical as it hides all the combinatorial aspects 
of the problem, and some of its calculations must be understood
in a purely formal way. 
Recently, a new purely combinatorial method for studying planar 
maps was developed, 
using so-called conjugacy classes of trees, allowing for a new
derivation of Tutte's results as well as some generalizations
(see \SCHrev\ for a review). In comparison to our present goal,
this approach was restricted to the enumeration of Eulerian planar 
maps, i.e. maps with only vertices of {\it even} valences.

In this paper, we give a combinatorial solution of the general problem
of enumeration of planar maps with arbitrary valences, generalizing the method of 
conjugated trees and elucidating the combinatorial structure
of the one-matrix model solution. The paper is organized as follows.
In Section 2, we introduce the quantities of interest and 
recall their expression as provided by the one-matrix model
(corresponding derivations are detailed in appendix A). 
In Section 3, we present our combinatorial proof. We first define 
appropriate decorated trees whose enumeration makes the
connection with the one-matrix model solution.  We then establish
bijections between classes of such trees and planar maps.
Particular cases are discussed in Section 4 and conclusions are
gathered in Section 5.

\newsec{Solution via matrix integral}

\subsec{Definitions}

Let us now come to the precise definition of the various 
generating functions that we wish to compute. 

Planar maps may display a large
number of internal symmetries making the counting quite subtle.
Indeed, in both matrix integral and combinatorial approaches,
maps are naturally counted with an inverse symmetry weight.
A convenient way to avoid this problem is to consider 
instead {\it rooted} maps, with a marked oriented edge which 
lifts up ambiguities due to symmetries.
We will denote by $E(t;\{g_k\}_{k\geq 1})$ the generating function for 
rooted planar maps.

Beside these ``closed diagrams", it will prove convenient to
also consider {\it one-leg diagrams}, namely planar maps with
a distinguished univalent vertex (the endpoint of the external leg). 
For these diagrams, no symmetry problem arises and we denote by 
$\Gamma_1(t;\{g_k\})$ their generating function with no
weight for the distinguished vertex.  

Finally, we will also consider the generating function 
$\Gamma_2(t;\{g_k\})$ for {\it two-leg diagrams} with two unweighted distinguished 
(as, say, incoming and outcoming) univalent vertices 
{\it adjacent to the same face}.   

\fig{Typical planar maps contributing respectively to the generating functions
$E$, $\Gamma_1$ and $\Gamma_2$, namely  (a) a rooted planar map with one 
regular univalent, one bivalent, three trivalent and two
tetravalent vertices; (b) a one-leg 
diagram with one regular univalent, one bivalent,
one trivalent, three tetravalent and one pentavalent vertices; 
(c) a two-leg diagram with one regular univalent, five trivalent and
two tetravalent vertices. In the first case, the rooted edge is marked
by an arrow. In the two latter, the distinguished univalent vertices 
attached to the external legs are marked by a cross. 
In (c), the two legs must lie in the same
face and are distinguished as
in- and out- coming via an orientation.}{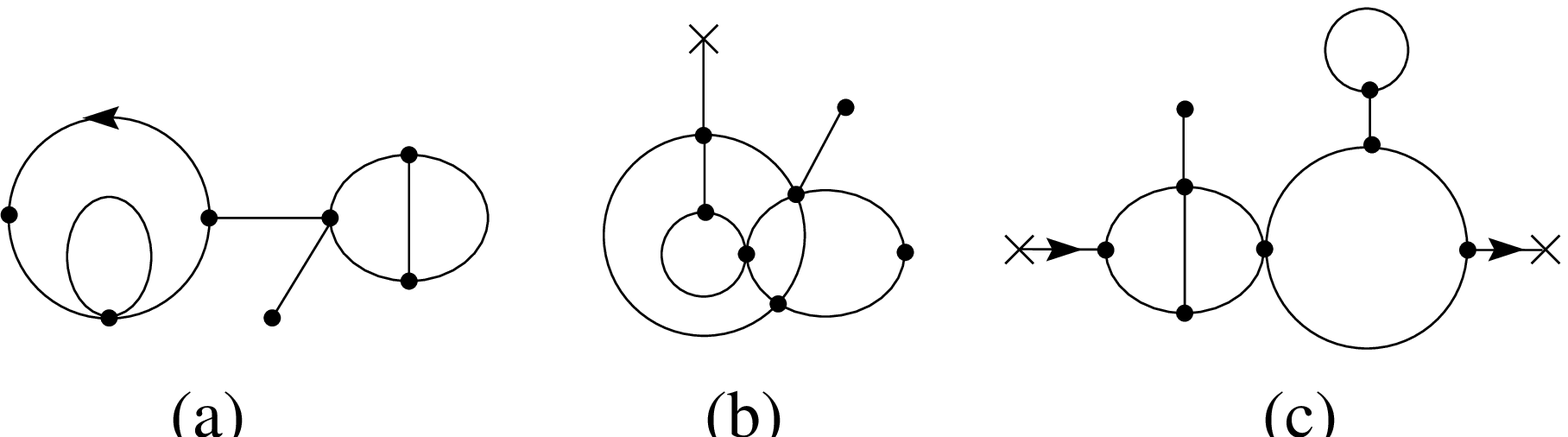}{12.cm}
\figlabel\map 

Examples of maps contributing to $E$, $\Gamma_1$ and $\Gamma_2$
are displayed in Fig. \map. Our convention to represent maps 
in the plane is to choose as the external face that on
the right of the marked edge for closed diagrams, and that
adjacent to the distinguished vertices for both one- and two-leg
diagrams.

When cutting the marked edge in a rooted closed map, we get either
two disconnected one-leg diagrams, or a two-leg diagram with legs
in the same face, excluding that formed by a single oriented edge.
This leads to the first general relation 
\eqn\relE{E={\Gamma_1^2+\Gamma_2-t\over t}}

\subsec{One-matrix model results}

Formally, all quantities of interest can be obtained as the
large $N$ limit of average values in a $N\times N$ one-matrix 
model with partition function
\eqn\partf{Z=\int dM \exp N {\rm Tr}\left(-{M^2\over 2 t}
+\sum_{k\geq 1}{g_k\over k} M^k \right)}
where $dM$ denotes the standard Haar (SU($N$) invariant) measure 
over hermitian matrices (see e.g. \DGZ\ and \EY\ for details). 

As shown in appendix A, the solution of the problem involves 
two functions $S(t;\{g_k\})$ and $R(t;\{g_k\})$ characterizing 
the eigenvalue distribution at large $N$.
Introducing the potential 
\eqn\pot{V(x)=\sum_{k\geq 1} {g_k\over k} x^k}
and the formal parametrization
\eqn\paraQ{Q(z)=z+S+R/z}
in terms of a dummy variable $z$, the functions $S$ and $R$
are implicitly determined by 
\eqn\SR{\eqalign{S &= t V'(Q)|_{z^0} \cr
R &= t+ t V'(Q)|_{z^{-1}} \cr}}
where $V'(Q)|_{z^m}$ denotes the coefficient of $z^m$ in $V'(Q)$
when viewed as a Laurent series in $z$.
The correct determination is fixed by 
$S(0;\{g_k\})=R(0;\{g_k\})=0$, allowing for an expansion as
a power series in $t$.

The generating functions for maps are then given
in terms of $S$ and $R$ through
\eqn\eqGamma{\eqalign{\Gamma_1 &= S-V'(Q)|_{z^{-2}} \cr
\Gamma_2 &= R-V'(Q)|_{z^{-3}}-\left(V'(Q)|_{z^{-2}}\right)^2\cr}}
and $E$ follows from Eq. \relE.

In principle, the above equations solve the enumeration problem.
In practice, the generic term of the series can be computed
explicitly in simple cases such as cubic maps ($g_k=0$ when 
$k\neq 3$), or Eulerian maps ($g_k=0$ for odd $k$)
as discussed in Section 4 below.
As an illustration, we list the first few terms in Eq. \SR\ by
expanding the potential \pot\ up to the quartic term
\eqn\SRexpand{\eqalign{S &= tg_1+tg_2S+tg_3(S^2+2R)+tg_4(S^3+6RS)+\cdots\cr
R &= t+tg_2R+tg_3(2RS)+tg_4(3S^2R+3R^2)+\cdots\cr}}
which can be solved order by order in $t$ with the initial conditions
$R=t+O(t^2)$, $S=tg_1+O(t^2)$.

In the next Section, we will recover precisely the same set
of equations through a purely combinatorial approach.

\newsec{Combinatorial Solution}

We now come to the core of the paper, namely the combinatorial
rederivation of the above equations \SR\ and \eqGamma.
We first show that the functions $S$ and $R$ are indeed 
generating functions of decorated ``blossom trees" generalizing
those introduced for the counting of Eulerian maps
in Ref. \SCH. These trees are to be closed into graphs
to recover planar maps. As in Ref. \SCH, bijections can be
established between planar maps and properly defined conjugacy
classes of these trees, which we will describe precisely.
We will show the equivalence to planar maps both in the case 
of one-leg and two-leg diagrams, allowing to recover all the
results of the matrix model described above.

\subsec{Rooted blossom trees}

\fig{A typical S-tree (S) and a typical R-tree (R). The leaves
are represented with empty arrows, buds correspond
to filled black arrows, while regular vertices are solid black dots.
The leaves carry a charge $+1$ while the buds carry a charge $-1$.
The S-tree has a total charge $0$ ($7$ leaves and $7$ buds),
while the R-tree has a total charge $1$ ($5$ leaves and $4$ buds).
Both have the property that any descendent subtree not reduced to
a bud has total charge $0$ or $1$.}{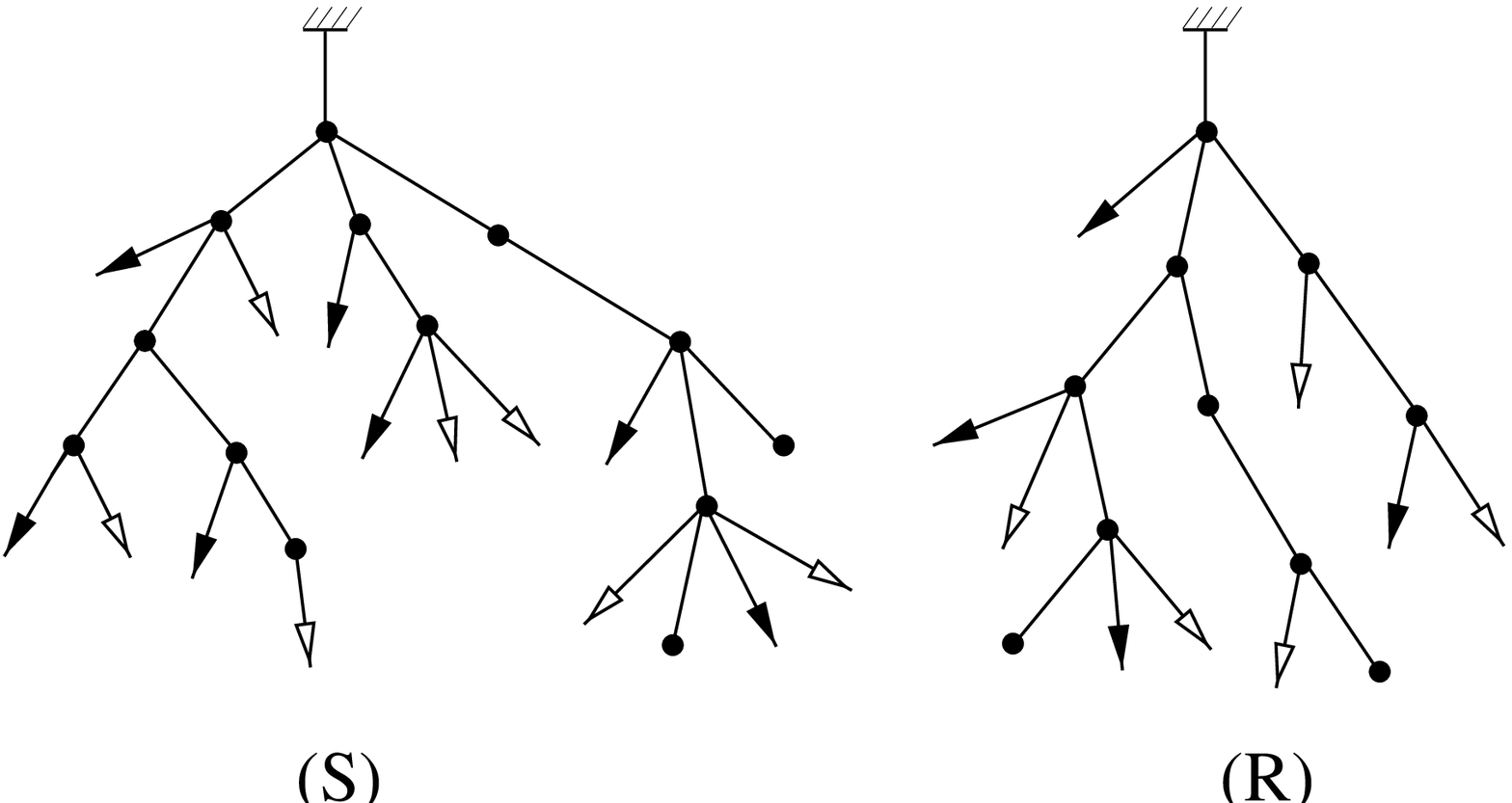}{12.cm}
\figlabel\SRsample

Let us now define rooted S- and R- {\it blossom trees}.
These are (finite)
rooted planar trees with three species of endpoints
that we will refer to as {\it leaves}, {\it buds}, and regular
univalent vertices. 
We assign a ``charge" $q=+1$ (resp. $q=-1$) to 
leaves (resp. buds), while regular univalent vertices remain
neutral ($q=0$). These trees are called S- (resp. R-) trees 
iff: (i) their total charge is $0$ (resp. $1$) and (ii) 
any descendent subtree not made of a single bud has
total charge $0$ or $1$. 

As an example, the smallest S-tree is made of a single regular 
univalent vertex attached to the root while
the smallest R-tree is made of a single leaf attached to the root. 
Typical S- and R- trees are represented in Fig. \SRsample.

\fig{Illustration of the recursive generation of a rooted S-tree
via the enumeration of all possible vertices of total charge $0$ attached to the root
for terms up to $g_4$ in the potential $V$. This parallels
the first line of the recursion \SRexpand\ by viewing the labels 
$S$ and $R$ as generating functions for S- and R-trees.}{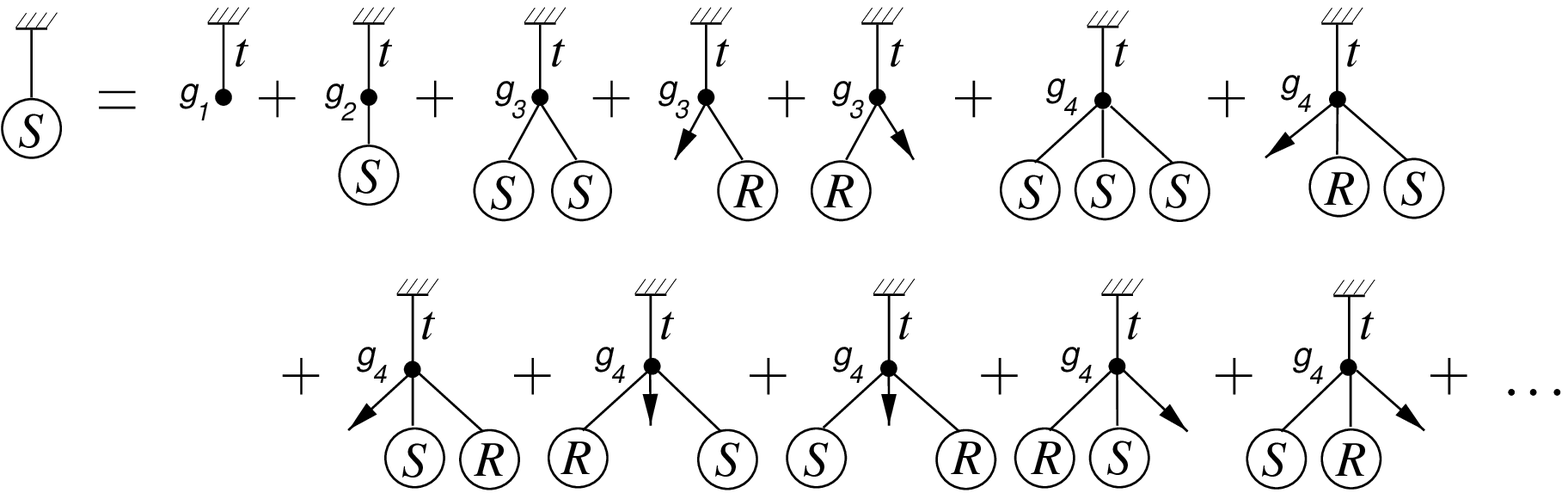}{12.cm}
\figlabel\Stree

\fig{The same illustration as in Fig. \Stree\ for a rooted R-tree
(with now vertices of total charge $1$), to
be paralleled with the second line of the recursion \SRexpand\ for
terms up to $g_4$ in the potential $V$.}{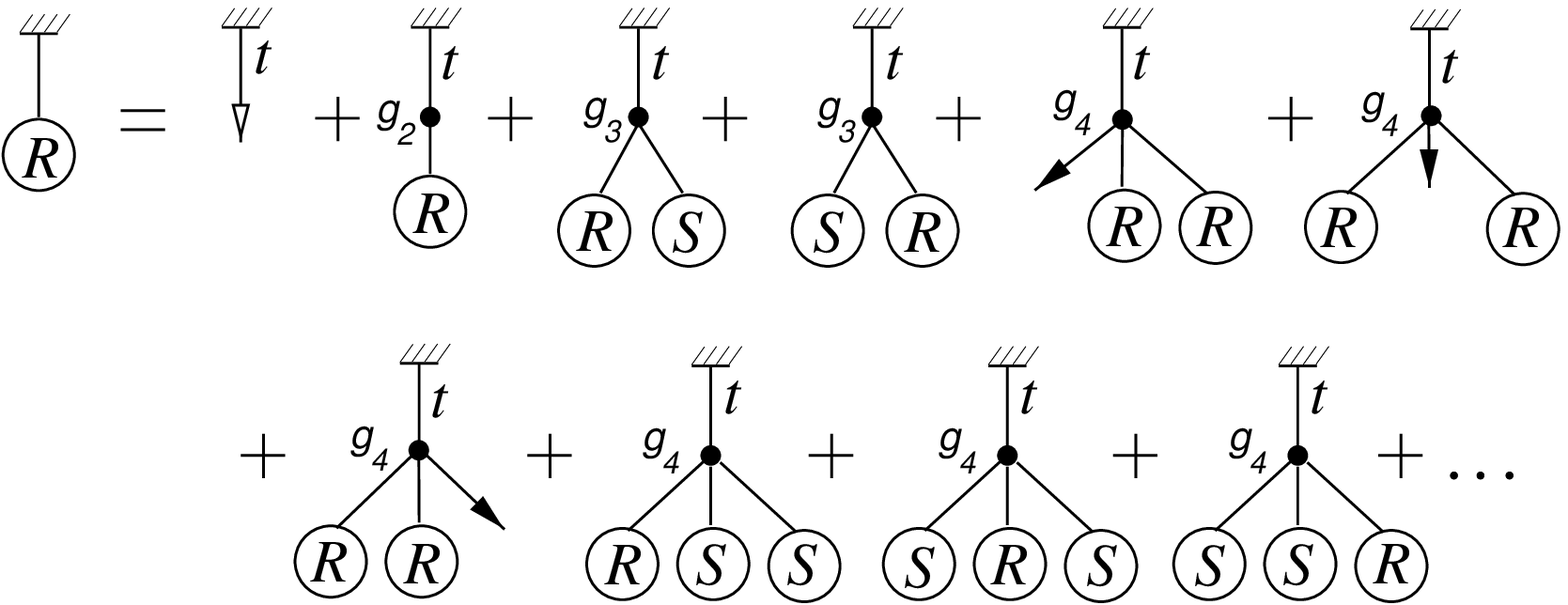}{10.cm}
\figlabel\Rtree

Obviously, any descendent subtree of an S- or R-tree not 
reduced to a bud is itself an S- or an R-tree.
This recursive property allows to interpret the functions $S$ (resp. $R$) of 
Eq. \SR\ as the generating functions for 
rooted S-trees (resp. R-trees) with a weight $g_k$ per $k$-valent 
vertex ($k\geq 1$) and a weight $t$ per edge not 
leading to a bud. First, we note that the dummy variable $z$ 
in Eq. \paraQ\ may be thought of as a fugacity per unit of charge. 
Viewing the three terms in $Q(z)=z+S+R/z$ as associated respectively 
to a bud ($q=-1$), an S-tree ($q=0$) and an R-tree ($q=+1$), 
the power of $1/z$ measures the total charge of any composite object. 
This allows to interpret $V'(Q)|_{z^{-m}}$ as generating all possible 
vertices with descendents being buds, S-trees and R-trees,
with a total charge $q=m$ and with weight $g_k$ per $k$-valent vertex
(i.e. with $(k-1)$ descendents). Expressions \SR\ 
for S-trees and R-trees follow by enumeration of all possible
configurations around the vertex attached to the root, with a total
charge of $0$ and $1$ respectively.  
The factor $t$ in front of $V'(Q)$
accounts for the weight of the root edge, while the additional $t$
term in $R$ stands for the germ of the recursion. These recursions
are depicted in Figs. \Stree\ and \Rtree, for the first 
few terms involving $1$, $2$, $3$ and $4$-valent vertices.

\subsec{Conjugated trees}

Starting with the above S- and R-trees, we may now wipe out 
the root by replacing it by a standard (unmarked) leaf, thus
increasing the total charge by one unit. This defines {\it unrooted}
S- and R-trees with total charge $1$ and $2$ respectively.
Two rooted S- (resp. R-) trees are said to belong to
the same {\it conjugacy class} iff they lead to the same unrooted
tree. 

The case of S-trees is simple in that reciprocally, given any
unrooted S-tree, we obtain all the corresponding rooted S-trees by picking 
{\it any} of its leaves and choosing it as the root. This
leads to a $n$-to-one correspondence between rooted and unrooted
S-trees with $n$ leaves (including the former root). 
The proof goes as follows. 
Recall first that cutting any edge not leading to a bud
in a rooted S-tree cuts out a descendent subtree of charge
$0$ or $1$.
Consequently, cutting the same edge in the corresponding unrooted
S-tree separates the tree into a piece of charge $0$ or $1$
and a complementary piece of charge $1$ or $0$ respectively,
as the total charge is $1$. This property, valid for any edge 
not leading to a bud, completely characterizes
the unrooted S-trees. Indeed, starting from an unrooted tree with
this property, we see that it has total charge $1$, and that 
therefore by replacing an arbitrary
leaf by a root, we end up with a rooted tree of total charge $0$ 
whose descendent subtrees not reduced to a bud have charge $0$ or 
$1$, the characterization of rooted S-trees. As the choice of leaf 
taken as a root is arbitrary among the $n$ leaves, this moreover 
shows that the conjugacy class is made of $n$ rooted S-trees,
by finally noting that the unrooted
S-trees have no accidental rotational symmetry since the
numbers of buds ($n-1$) and leaves ($n$) are coprime.    
This completes the abovementioned $n$-to-one correspondence.
A last remark is in order: cutting any edge not leading to a bud of
an unrooted S-tree separates the tree into two rooted trees, one
being a rooted S-tree and the other a rooted R-tree. Indeed 
these trees have respective charges $0$ and $1$ and 
any of their descendent subtrees not made of a single bud
also have charge $0$ or $1$ as pieces of the unrooted S-tree.

\fig{The core of a typical unrooted R-tree. We have indicated
by double-lines the edges ``of type $0$-$2$" to be cut in the
core construction process. The circled pieces
are the maximal S-subtrees (of charge $0$) attached to the core 
(of charge $2$), represented
in thick lines. In this particular example, the core has
three leaves and one bud.}{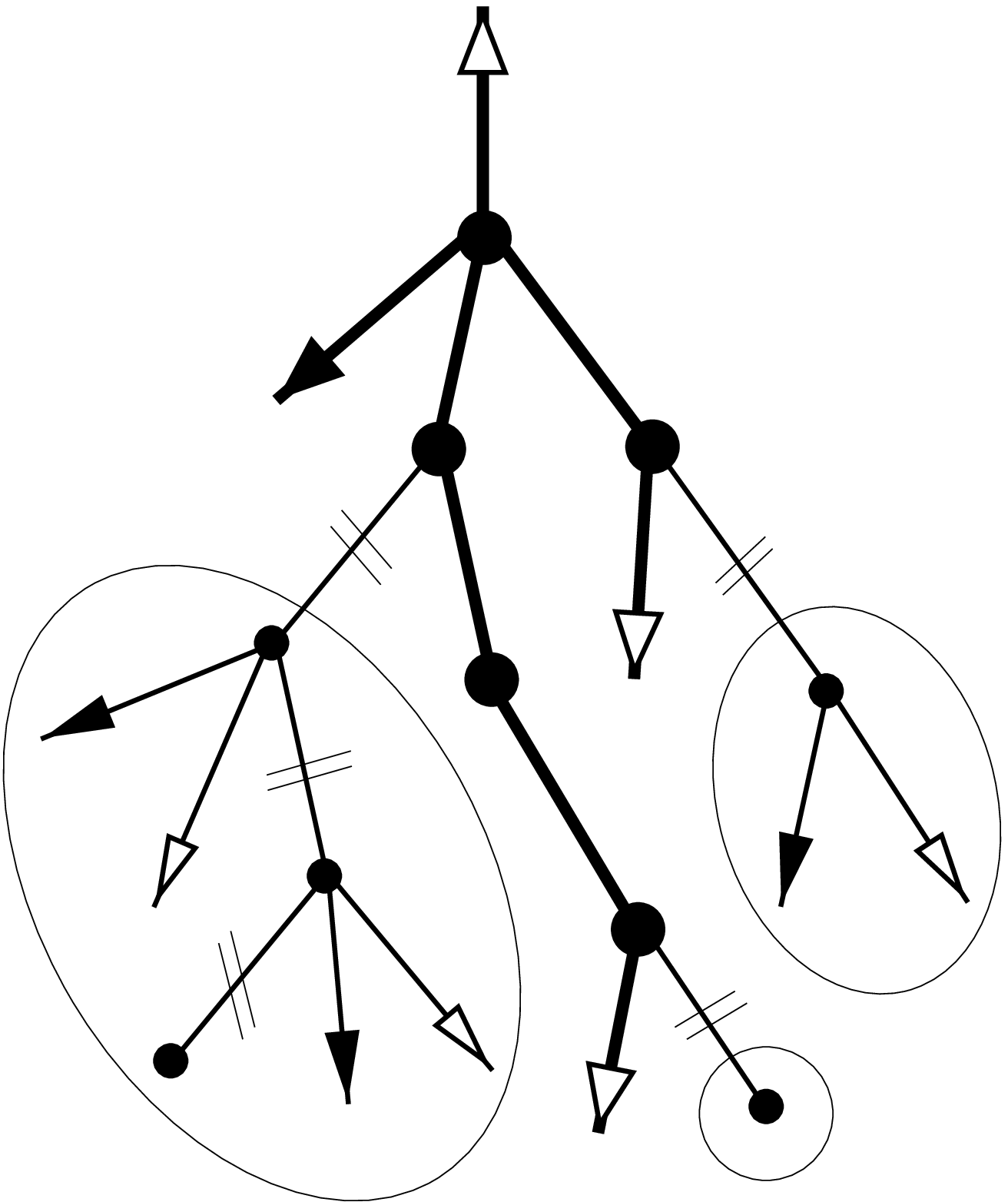}{5.cm}
\figlabel\Runrooted

The case of R-trees is more subtle. Indeed, it is no longer true that
picking as a root any leaf of an unrooted R-tree leads to a rooted R-tree. 
In order to characterize the subset of {\it admissible} leaves
(leading effectively to rooted R-trees), we first define the {\it core} of
an unrooted R-tree 
by the following procedure: (i) we mark all the inner edges which
separate the tree into two pieces of respective charge $0$ and $2$, 
(ii) we remove all these edges and (iii) we define the core as the
only connected component with total charge $2$. This is illustrated
in Fig. \Runrooted.
To show the existence and uniqueness of the core, we consider
an arbitrary admissible leaf and the corresponding rooted R-tree.
Due to charge constraints, an edge is marked at step (i) iff the descendent
subtree originating from it is an S-tree. Such a subtree is then amputated from
its own S-subtrees, splitting eventually into connected components all of
charge $0$. The connected component containing the root is the only one
not obtained in this way, hence it has charge $2$. 
Note also that all trees attached to the core are rooted
S-trees attached by their root. We call them {\it maximal} S-subtrees. 
As our procedure
was defined independently of the admissible root at hand, it follows that
all admissible leaves belong to the core. 
Conversely, replacing an arbitrary leaf in the core by a root,
we get a rooted tree such that any descendent subtree not made of a single bud
either is contained in an S-subtree, or originates from an edge in the
core not marked at step (i): in both cases its total charge is $0$ or $1$
as wanted. 
To conclude, the conjugacy class associated with an unrooted R-tree
is made of all trees rooted at leaves of the core. As opposed to the S-case, 
care must be taken with a possible two-fold
rotational symmetry (half turn) which is the only one compatible 
with the fact that there are two more leaves than buds in an unrooted 
R-tree.

\fig{The core of a typical rooted R-tree as obtained from the
semi-developed representation using only the vertices of Fig. \Rtree.
Replacing the ``leaves" marked with a label $S$
by arbitrary rooted S-trees, we generate all rooted R-trees
sharing the same rooted core.}{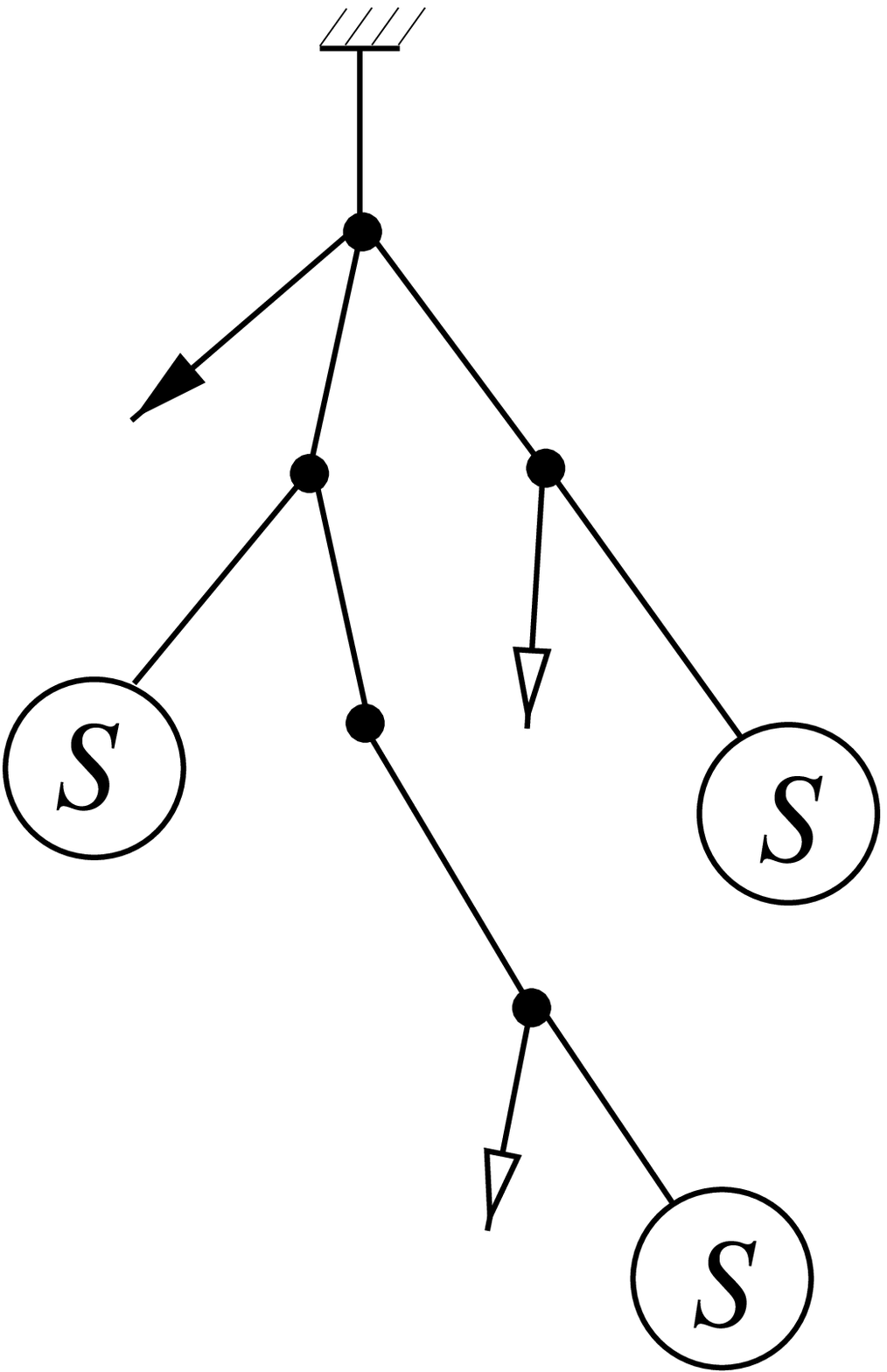}{4.cm}
\figlabel\Rcore

To conclude this section, let us note that a simple way to generate
the core of R-trees is to use a semi-developed representation
of rooted R-trees, namely by only using the vertices of Fig. \Rtree\ 
(second line of Eq. \SR) and keeping S as a label, as illustrated in Fig. \Rcore.

\subsec{Enumeration of one-leg diagrams:}

Let us now show that one-leg diagrams are in one-to-one correspondence
with unrooted S-trees.

\fig{A sample unrooted S-tree is closed into a one-leg diagram in a unique way,
by iteratively matching each bud to the closest available leaf 
in counterclockwise direction. The resulting edges are indicated in thin 
solid lines, and form a system of non-intersecting arches around 
to the tree. We have also indicated for each bud and leaf
the corresponding depth, namely that of the arch in the arch system,
starting with depth $1$ for arches adjacent to the external face.
The unmatched vertex receives the depth $0$ and serves as the
distinguished endpoint of the one-leg diagram.}{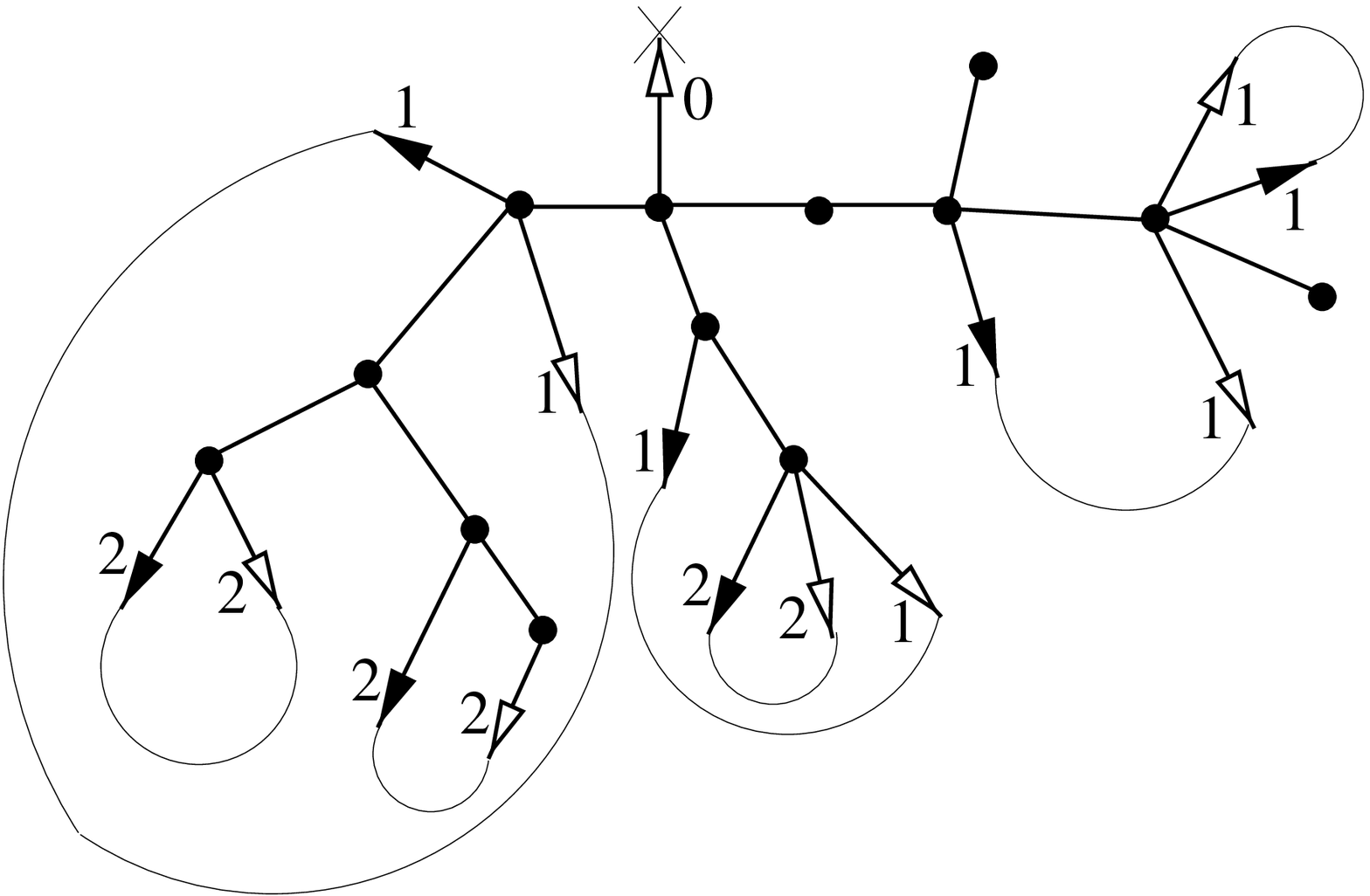}{10.cm}
\figlabel\StoGamma

Starting from an unrooted S-tree, we build a map by matching buds
and leaves as in Ref. \SCH,  by connecting iteratively each bud
to the closest available leaf in {\it counterclockwise} direction as
shown in Fig. \StoGamma, in such a way 
that the resulting graph is planar (i.e. with no intersection of edges). 
Such connected 
bud-leaf pairs are replaced by regular edges. As the total charge
of $1$ counts the number of leaves minus that of buds, the matching procedure
leaves exactly one unmatched leaf which we replace by a distinguished
univalent vertex. The net result is a one-leg diagram.

\fig{A sample one-leg diagram is cut into an unrooted S-tree by visiting
edges in counterclockwise direction around the diagram and iteratively cutting those
which do not disconnect the remaining diagram, until a tree is
obtained. The cut edges are indicated 
by a double-line, they are to be replaced by a bud, followed by a leaf. 
We have also indicated the depth of the corresponding cut edges,
which will translate into the depths of buds and leaves of the S-tree. 
These depths correspond to the number of visits to the distinguished
endpoint before the edge is cut.}{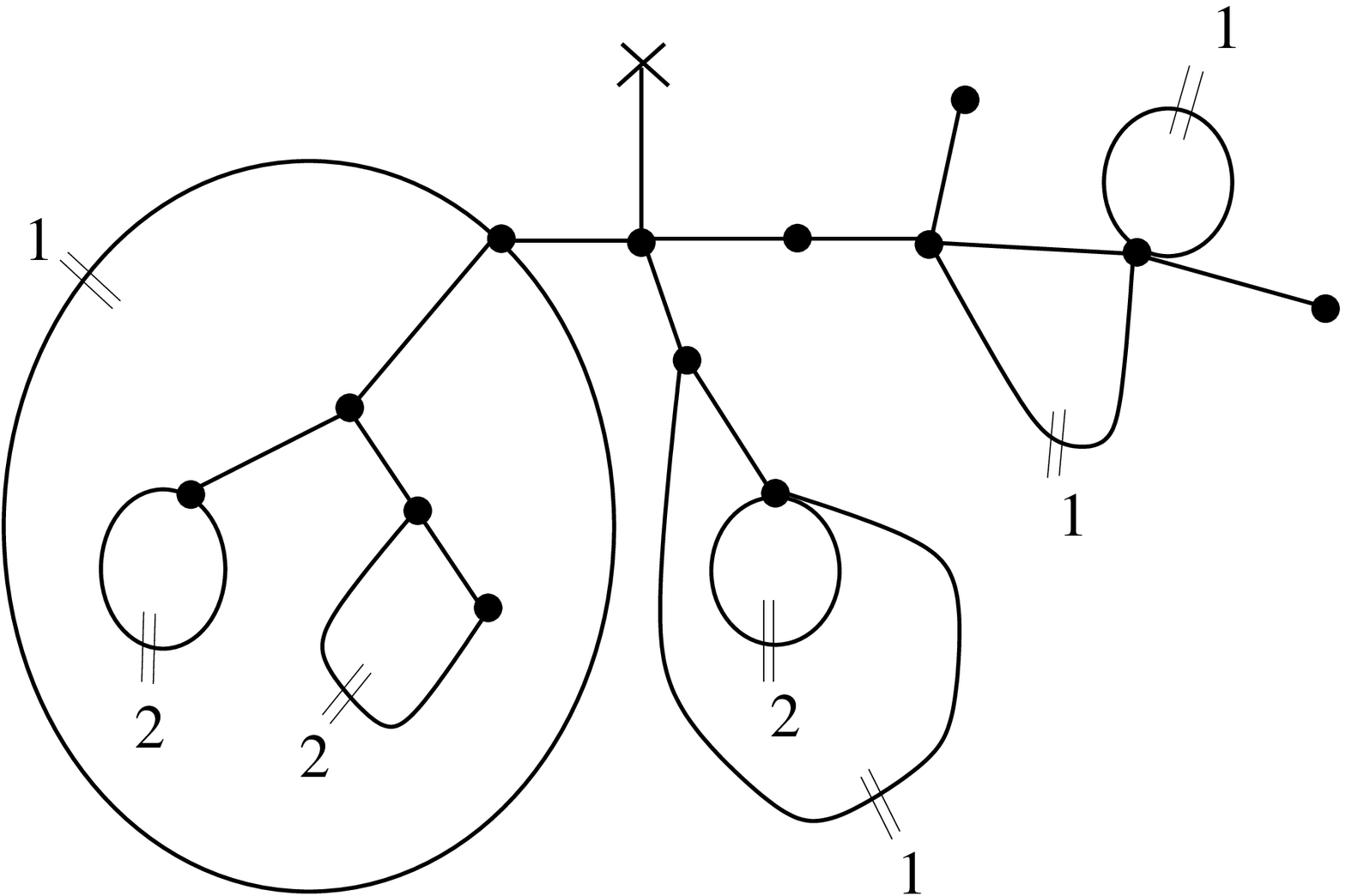}{9.cm}
\figlabel\GammatoS

Conversely, starting from a one-leg diagram, we recover an unrooted 
S-tree by an inverse algorithm similar to that of Ref. \SCH. Starting from the 
distinguished endpoint, we successively visit all edges adjacent to the 
external face in {\it counterclockwise} direction as shown in Fig. \GammatoS.
At each step, the edge is cut iff the cutting does not disconnect the diagram.
In the cutting procedure, the first half of the edge is replaced by a bud and 
the second half by a leaf. After one turn, the
external face has been merged with all its adjacent faces. We then iterate the procedure
until all faces have been merged. Replacing the distinguished endpoint
by a leaf, the resulting connected diagram is 
an unrooted blossom tree with buds, leaves and regular vertices,
and with total charge $1$. 

\fig{The tree obtained by cutting a one-leg diagram is an unrooted S-tree. 
Any of its inner edges, $e$, separates the tree into two pieces $T_1$ and $T_2$,
say with the distinguished endpoint in $T_1$. We have represented the
cut bud-leaf pairs formerly connecting $T_1$ to $T_2$. The 
cutting procedure has successively cut these pairs in counterclockwise
order around the edge $e$. In our example,
there is one more bud-leaf pair on the left than on the right
of $e$ (odd number of passings). 
We could also have the same number
of bud-leaf pairs on both sides (even number of passings). 
Collecting the total charge in $T_1$ and $T_2$ we see that in the case
of the figure $T_1$ has charge $0$ while $T_2$ has charge $1$.
In the case of an even number of passings, $T_1$ would have charge $1$
and $T_2$ charge $0$. This is the charge characterization of an 
unrooted S-tree.}{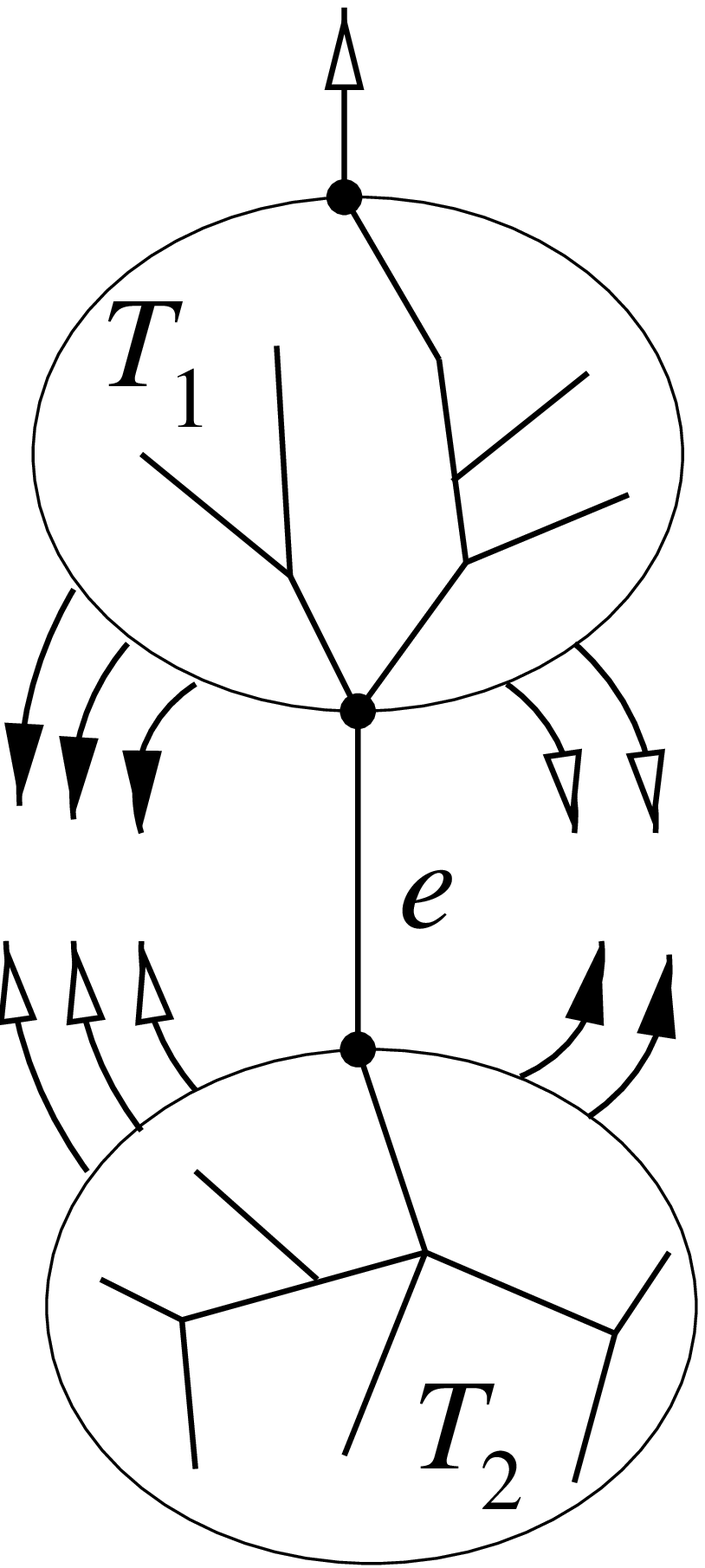}{3.cm}
\figlabel\itsanS

Let us now show that this tree actually is an S-tree by
considering one of its inner edges (connecting regular vertices),
which we denote by $e$ (see Fig. \itsanS). 
This edge separates the tree into two
pieces, say $T_1$ and $T_2$, which contain all the vertices of
the original diagram. We choose for $T_1$ the piece containing 
the distinguished endpoint. In the cutting procedure, $e$ is 
not visited until all the other edges of the original diagram 
connecting a vertex of $T_1$ to one of $T_2$ have been visited once 
and cut into a bud-leaf pair. More precisely, these edges are replaced
by a bud in $T_1$ and a leaf in $T_2$ when passing from $T_1$ to $T_2$
and {\it vice versa}. As the passings from $T_1$ to $T_2$ and
from $T_2$ to $T_1$ alternate, these buds and leaves contribute by 
a net charge of $0$ in both $T_1$ and $T_2$ if the number of passings
is even, while they contribute by a net charge of $-1$ in $T_1$
and $+1$ in $T_2$ if the number is odd. Adding the charge $+1$ of the
distinguished vertex changed into a leaf and noting that all 
other edges not connecting $T_1$ to $T_2$ 
have neutral contributions,
we end up with two pieces of respective charges $0$ and $1$, 
the characterization of unrooted S-trees.

To further establish the bijection, it remains to prove that
the two algorithms described above are inverse of each other.
This is best seen by introducing the notion of {\it depth} of 
buds and leaves within an unrooted S-tree. Starting from a 
one-leg diagram and applying the above cutting algorithm, we
may associate to each cut edge a natural ``depth" 
$d=1,2,3,...$ by the number
of visits to the distinguished vertex before this edge is cut.
The same depth is attached to the corresponding bud and leaf,
and finally the extra leaf replacing the distinguished vertex
is given the depth $0$. 
We have indicated the depths of buds and leaves in Figs. \StoGamma\ and
\GammatoS\ for illustration.
Remarkably, the depth is a notion 
intrinsic to the resulting unrooted S-tree. Indeed,
in the bud-leaf matching procedure of an unrooted S-tree, the 
created edges form a system of arches around the tree and
the depth of the buds and leaves is nothing but that of the
corresponding arches, starting with depth $1$ for 
external arches, and moreover associating the depth $0$ 
to the unmatched leaf. 
This system of arches may also be constructed now as
leaf-bud pairs of {\it decreasing} depth as follows. 
Starting from the depth $0$ leaf we proceed {\it clockwise} around
the tree and connect the first encountered leaf-bud pair of maximal
depth, say $k$, and continue to connect leaves and buds of depth $k-\ell+1$ 
after $\ell$  visits to the depth $0$ leaf.   
This alternative closing procedure is the exact inverse
of the cutting algorithm for one-leg diagrams.

As a direct consequence of the above bijection, the enumeration
of one-leg diagrams is equivalent to that of unrooted S-trees.
The latter is very simply performed by noting that there
is one more leaf than bud in such a tree.
The corresponding generating function is therefore equal to
the difference between that of 
unrooted S-trees with a marked leaf 
and that of unrooted S-trees with a marked bud.
The generating function of unrooted S-trees with a marked leaf
is nothing but $S$ as each of these leaves may be taken as a root.
On the other hand, the generating function for unrooted S-trees 
with a marked bud is computed 
by noting that these trees are in one-to-one correspondence
with rooted trees of total charge $2$ whose descendent subtrees are
buds, S- or R-trees, as is easily seen by replacing the marked bud 
by a neutral root. The generating function for unrooted S-trees 
with a marked bud therefore reads  
$V'(Q)\vert_{z^{-2}}$, leading to $\Gamma_1=S-V'(Q)\vert_{z^{-2}}$
as in \eqGamma. QED.

\subsec{Enumeration of two-leg diagrams:}

\fig{A typical two-leg diagram (a) is cut into an unrooted
R-tree (b) by first applying the cutting algorithm (cut edges are indicated
by parallel thin lines in (a)) and then replacing the in- and
out- coming endpoints by leaves. We have indicated the depths
of the leaves and buds in the R-tree (b), as well as maximal S-subtrees
(circled pieces) and the core to which they are attached (thickened edges).
Note that the outgoing (depth $0$) leaf is in the core, while the incoming
one is not, as it lies in a maximal S-subtree. This gives an example
of an unrooted R-tree in the set R$_1$, i.e. with exactly one depth $0$
leaf in the core.}{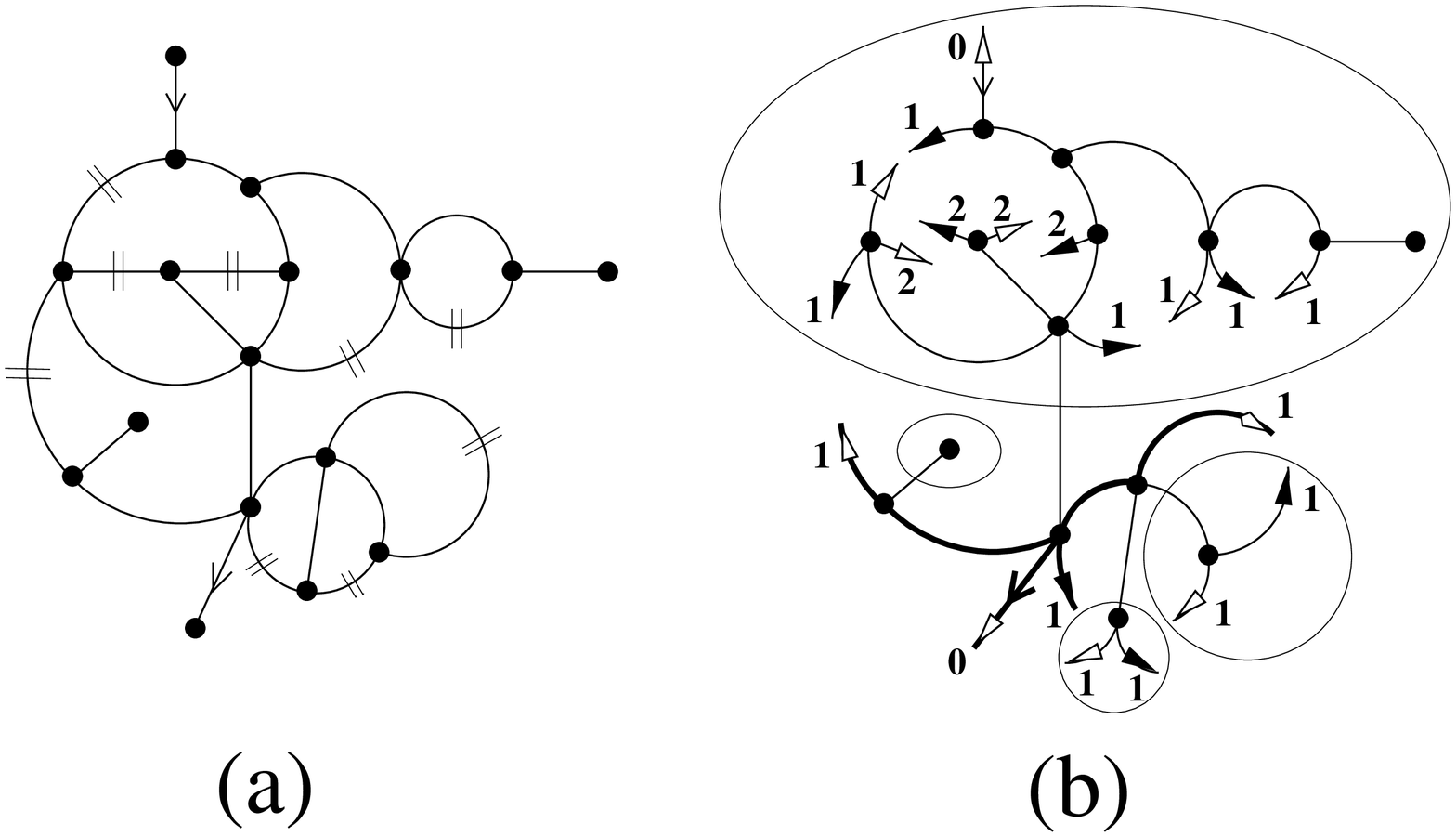}{12.cm}
\figlabel\gatwo

Let us now look for a similar equivalence in the case of two-leg 
diagrams. We will show that two-leg diagrams
are in one-to-one correspondence with unrooted R-trees {\it with
a marked leaf of depth $0$ in the core}.

We apply the same cutting procedure as before to two-leg diagrams
(see Fig. \gatwo\ (a))
starting from the {\it incoming} distinguished endpoint and finally
replacing the two distinguished endpoints by leaves. The resulting 
tree (see Fig. \gatwo\ (b)) has total charge $2$. 
Let us now show that: (i) it is an unrooted
R-tree and (ii) the originally {\it outcoming} distinguished endpoint
is a leaf belonging to the core.
This is easily seen by noting that in the cutting procedure,
the outcoming endpoint is treated as a regular univalent vertex.
Therefore, viewing the original two-leg diagram as a one-leg
diagram with an extra marked regular univalent vertex, we may
apply the result of Section 3.3 showing that the resulting tree
is an unrooted S-tree with this same marked regular univalent
vertex. Cutting the edge leading to this vertex leaves us 
with two pieces, one rooted S-tree and one rooted R-tree.
Obviously, the S-tree is the piece made of the univalent
vertex attached to a root. Therefore, the other piece is
a rooted R-tree. This proves that picking the outcoming endpoint
as a root leads to an R-tree showing both (i) and (ii) as
the admissible roots are the leaves of the core.

Up to this point, we have never used the fact that
the two endpoints of the diagram both lie in the external face.
In fact, the above construction shows that $R$ can be understood
as the generating function $\Gamma_{1,1}$ of two-leg diagrams
whose legs do not have to lie in the same face. This
is also proved in appendix A in the matrix language.
The requirement that the two endpoints belong to the external face
implies a further restriction on the unrooted R-tree obtained
through the cutting procedure, namely that 
the leaf replacing the outcoming endpoint
has depth $0$, as illustrated in Fig. \gatwo\ (b). 
As before, we indeed have an intrinsic notion of depth
for unrooted R-trees, defined through the same bud-leaf 
counterclockwise matching algorithm. 
For unrooted R-trees, it now leads to two arch systems separated
by two unmatched leaves. These two leaves are assigned a depth $0$
while all the other buds or leaves are assigned the depth
of the corresponding arch. For two-leg diagrams, the external 
legs naturally separate the edges to be cut into two independent
arch system, showing that, when replaced by leaves, the incoming
and outcoming endpoints both have depth $0$ (see Fig. \gatwo\ (b)). 
Finally, the inverse of the above cutting procedure is easily identified
as before with the suitable alternative matching algorithm connecting
leaves and buds clockwise in decreasing depth order. 

The enumeration of two-leg diagrams is finally reduced to that
of unrooted R-trees with a marked leaf of depth $0$ in the core.
As a preliminary remark, note that there are exactly two leaves
of depth $0$ in an unrooted R-tree, which may or may not belong
to the core. This suggests to classify the unrooted R-trees 
into three subsets R$_0$, R$_1$ and R$_2$ according to 
the number $0$, $1$ or $2$ of depth $0$ leaves in the core.
For instance, the unrooted R-tree of Fig. \gatwo\ (b)
belongs to R$_1$, as the incoming (depth $0$) leaf lies in a maximal
S-subtree (circled in the figure), while the outcoming one is in the core
(represented by thickened edges in the figure).
Denoting by $R_0$, $R_1$ and $R_2$ the corresponding generating
functions, we have 
\eqn\bigam{\Gamma_2=R_1+2 R_2=2{\tilde R}-(R_1+2R_0)}
where ${\tilde R}\equiv R_0+R_1+R_2$ is the generating function 
for unrooted R-trees. 

${\tilde R}$ can be computed as before by
noting that there are exactly two more leaves than buds {\it in
the core}. Hence $2{\tilde R}$ is the difference between the generating
function for unrooted R-trees with a marked leaf in the core,
and that for unrooted R-trees with a marked bud in the core.
The former is nothing but $R$ according to Section 3.2.
On the other hand, the generating function for unrooted R-trees 
with a marked bud in the core is computed 
by noting that these trees are in one-to-one correspondence
with rooted trees of total charge $3$ whose descendent subtrees are
buds, S- or R-trees. This is seen again by replacing the marked bud 
by a neutral root and checking that all proper descendent subtrees
not reduced to a bud have charge $0$ or $1$. Indeed
such a descendent subtree either is itself a descendent of an S-subtree,
hence has charge $0$ or $1$, or originates from an edge of the core,
hence has charge $1$. The generating function for unrooted R-trees 
with a marked bud in the core therefore reads  
$V'(Q)\vert_{z^{-3}}$, leading to $2{\tilde R}=R-V'(Q)\vert_{z^{-3}}$.
Remark that the accidental two-fold symmetry of unrooted R-trees 
is properly accounted for in ${\tilde R}$ in which symmetric contributions
(which contribute only once to $R$ and $V'(Q)\vert_{z^{-3}}$) receive
a weight $1/2$. 
 
The remaining term in Eq. \bigam\ can be computed with the result 
$(R_1+2R_0)=(V'(Q)\vert_{z^{-2}})^2$. The proof is slightly tedious
and is detailed in appendix B.

This finally allows to express Eq. \bigam\ in the form of Eq. \eqGamma. QED.

\newsec{Discussion} 

\subsec{Cubic maps}

The first non-trivial case of planar maps is that of {\it cubic}
maps, namely maps whose regular vertices are all trivalent
(closed planar cubic maps are dual to triangulations of the sphere). 
These correspond
in our notations to taking $g_k =g \delta_{k,3}$, in which
case $V'(Q)=g Q^2$, and the fundamental relations \SR\ read
\eqn\srcub{ \eqalign{
S&=gt(S^2+2 R)\cr
R&=t+2 g t R S \cr}}
while the one- and two-leg diagram generating functions read
\eqn\onetwocub{\eqalign{
\Gamma_1&=S-g R^2\cr
\Gamma_2&=R-g^2 R^4\cr}}
Note that S- and R-trees are now binary trees, as they originate from cubic maps.
In this particular case, we also have the following
additional relation
\eqn\specub{ \Gamma_1= gt (\Gamma_2+\Gamma_1^2) }
obtained by cutting out from any one-leg diagram the leg and
the trivalent vertex connected to it: the remaining part is made
either of a connected two-leg diagram (term in $\Gamma_2$) or of
two disconnected one-leg diagrams (contributing to a term
in $\Gamma_1^2$).
Alternatively, the relation \onetwocub\ is also easily proved
by use of Eqs. \srcub\ and \onetwocub.

\fig{In the cubic map case, the core of a typical unrooted R-tree
is a chain with a leaf at each end and attached maximal
S-subtrees on either side.}{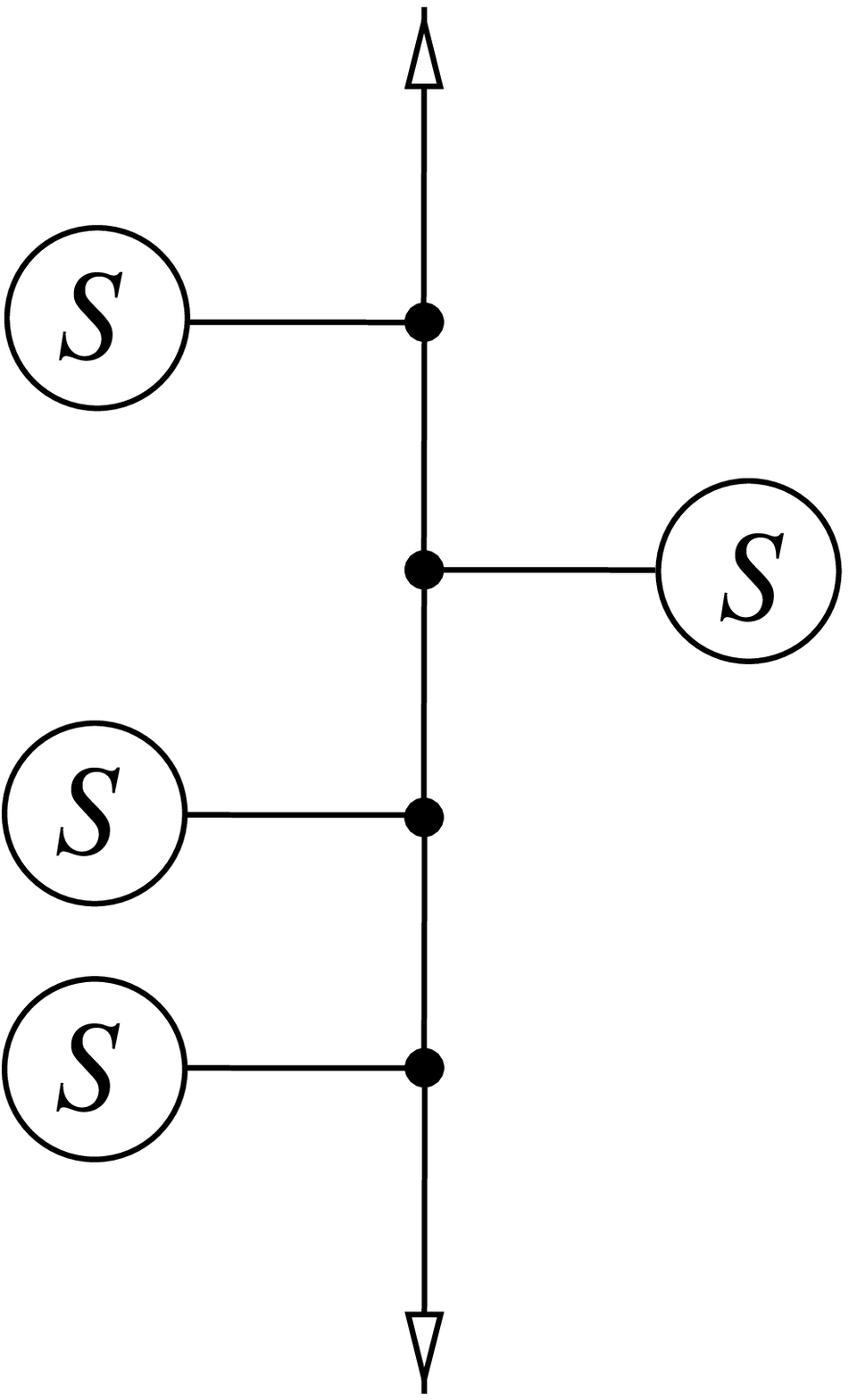}{4.cm}
\figlabel\corcub

Another drastic simplification is that the core of any unrooted
R-tree is a simple chain joining two leaves, with no buds and
with attached maximal S-trees (see Fig. \corcub). The absence of buds
is a consequence of $V'(Q)|_{z^{-3}}=0$ in this case, leading
moreover to $2{\tilde R}=R$, as there are two admissible leaves that
can serve as roots.

The actual numbers of S- and R-trees as well as one- and two-leg diagrams
with fixed number of edges are
obtained by first eliminating $R$ from Eq. \srcub, and then applying the 
Lagrange inversion formula. Introducing the rescaled functions and variables 
$\sigma=g t S$, $\rho=(gt)^2 R$, $\theta=g^2t^3$, we get
\eqn\eqsig{\eqalign{2\theta &= 
\sigma(1-\sigma)(1-2 \sigma) \cr
\rho&={\sigma(1-\sigma)\over 2} \cr} }
Denoting by $\varphi(\sigma)=\sigma(1-\sigma)(1-2 \sigma)$, we may now
express the series expansion in $\theta$ of any function $h(\sigma)$
(with $h(0)=0$) through the Lagrange inversion formula, a direct consequence of
the Cauchy formula:
\eqn\lag{\eqalign{
h(\sigma)&=\oint {ds \over 2i\pi} {h(s) \varphi'(s) \over \varphi(s)-\varphi(\sigma)}\cr
&=\sum_{n=1}^\infty {(2\theta)^n\over n} \oint {ds \over 2i\pi}{h'(s)\over \varphi(s)^{n}}\cr}}
where we have used the first line of \eqsig\ and integrated by parts. 
Upon picking respectively $h(x)=x$ and $h(x)=x(1-x)/2$, we get the series coefficients
$\sigma_n$ and $\rho_n$
for $S=(gt)^{-1}\sum_{n=1}^\infty \sigma_n \theta^n$
and $R=(gt)^{-2} \sum_{n=1}^\infty \rho_n \theta^n$ as contour integrals, easily
calculated as $\sigma_1=2$, $\rho_1=1$ and for $n\geq 2$:
\eqn\resrhosig{\eqalign{
\sigma_n&={2^n\over n} \oint {ds \over 2 i \pi s^{n}} {1\over (1-s)^n(1-2s)^n} \cr
&={2^n\over n}\sum_{k=0}^n 2^k {2n-k-2\choose n-1} {n+k-1\choose n-1} \cr
&={2^{2n-1} \over n!} (n+1)(n+3)...(3n-5)(3n-3)\cr
\rho_n&={2^{n-1}\over n} \oint {ds \over 2 i \pi s^{n}} {1\over (1-s)^n(1-2s)^{n-1}}\cr
&={2^{n-1}\over n}\sum_{k=0}^n 2^k {2n-k-2\choose n-1} {n+k-2\choose n-2} \cr
&={2^{2n-2} \over n!} n(n+2)(n+4)... (3n-6)(3n-4)\cr }}
for the numbers $\sigma_n,\rho_n$ of rooted S- and R-trees with $n$ 
leaves.
 
Similarly, we get expressions for the numbers of one- (resp. two-) leg diagrams
with $3n-1$ (resp. $3n-2$ edges), denoted by $\gamma_n^{(1)}$ 
(resp. $\gamma_n^{(2)}$) by expressing the relevant generating
functions in terms of $\sigma$ only, with the result $\gamma_1^{(1)}=1$,
$\gamma_1^{(2)}=1$ and for $n\geq 2$:
\eqn\resgaonto{\eqalign{
\gamma_n^{(1)}&= {2^{2n-1} \over (n+1)!} (n+1)(n+3)...(3n-5)(3n-3) \cr
\gamma_n^{(2)}&= {3\over n+2} {2^{2n-2} \over n!} n(n+2)(n+4)... (3n-6)(3n-4)\cr}} 
As a direct consequence of the S-tree conjugacy, we have $\gamma_n^{(1)}=\sigma_n/(n+1)$, as
a corresponding unrooted S-tree has exactly $n+1$ leaves which all may serve
as a root in the conjugacy class. 
We note a slightly less trivial relation between the numbers of two-leg diagrams and R-trees:
$\gamma_n^{(2)}=3 \rho_n/(n+2)$ which still awaits a good combinatorial interpretation. 

Finally, we get the number $e_n$ of rooted cubic maps with $3n$ edges by 
substituting Eq. \specub\ into Eq. \relE, leading to $E=\Gamma_1/(gt^2) -1$, 
and therefore
\eqn\nume{ e_n=\gamma_{n+1}^{(1)}={2^{2n+1} \over (n+2)!} (n+2)(n+4)(n+6)...(3n-2)(3n) }
for all $n\geq 1$.

\fig{The local environment in an unrooted S-tree in the cubic
case around (a) a leaf, (b) a bud, (c) a vertex with no bud, and
(d) an edge not leading to a bud.}{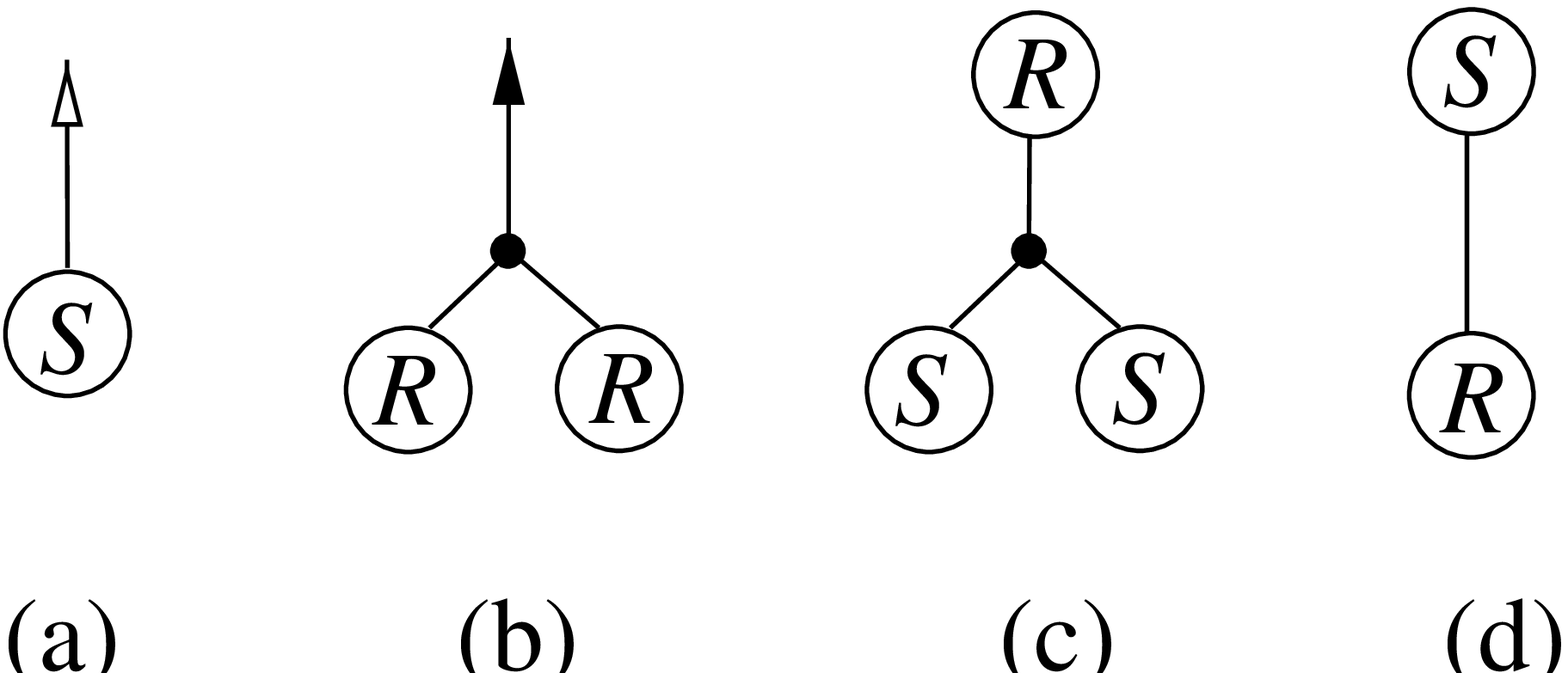}{10.cm}
\figlabel\orig

Introducing the rescaled one-leg diagram generating function 
$G_1=gt \Gamma_1$, we easily get the following differential
equations 
\eqn\difga{\eqalign{
\left(\theta {d\ \over d\theta} +1\right)G_1&=\sigma\cr
\theta {d\ \over d\theta} G_1&={\rho^2\over \theta} \cr
\left(\theta {d\ \over d\theta} -1\right)G_1&={\rho \sigma^2\over \theta}\cr
\left(3\theta {d\ \over d\theta} -1\right)G_1&={\rho\sigma\over \theta}\cr
}}
obtained by marking in an unrooted S-tree respectively a leaf, a bud, a trivalent vertex 
with no bud and an edge, in respective numbers $n+1$, $n$, $n-1$, $3n-1$.
The combinatorial origin of these relations is depicted in Fig. \orig.

\subsec{Eulerian maps}
 
The case of Eulerian maps, namely of diagrams with only regular vertices of 
{\it even valences} was solved by Tutte \TUTthree, and takes a much simpler form
than the general case. Eulerian maps may be recovered here by
demanding that $V$ be an even function, namely that all $g_{2k-1}=0$, $k=1,2,...$ 
The matrix model solution (see appendix A) shows in that case that $S=0$
as a consequence of the symmetry $x\to -x$ of the planar eigenvalue
density. From the present combinatorial point of view, it is easy to see that
the recursion relations of Figs. \Stree\ and \Rtree\ lead to infinite
rooted trees unless we decide that there are no rooted S-trees at all. In other words,
our finiteness requirement forces us to take the trivial generating
function $S=0$.

With this simplification, we are left with only rooted R-trees, which become 
the fundamental combinatorial objects, now characterized by the property
that (i) their total charge is $1$ and (ii) all their descendent subtrees not
reduced to a bud have charge $1$ as well (thus are themselves R-trees).
These trees are those which have been considered in Ref. \SCH\ in the original
connection between Eulerian maps and conjugated trees. 
Moreover, in the absence of S-trees, it is easy to see that the
core of an unrooted R-tree is actually the whole tree itself, with all
the leaves as admissible roots.
The notion of conjugacy therefore becomes much simpler and coincides with that 
used in Ref. \SCH. 


\newsec{Conclusion}

In this paper, we have elucidated the combinatorial structure
of the one-matrix model solution to the problem of enumerating
planar maps with prescribed vertex degrees. Our construction generalizes
that of Ref. \SCH\ for Eulerian maps using two types of trees instead
of one and a refined notion of conjugacy of trees. The need for 
these two types of trees is inherent to the absence of (twofold) 
symmetry in arbitrary planar graphs as opposed to Eulerian ones
(face bicolorability) which can be traced back to the absence
of $\IZ_2$ symmetry in the matrix model. 

Here we have considered arbitrary graphs with no restrictions
on reducibility. {\it Irreducible maps} were enumerated in some 
particular cases by Tutte \TUTone\ \TUTtwo\ \TUTfour\ and a connection with conjugated 
trees was obtained recently in Ref. \PoSc. In the matrix integral language, 
the construction of ($p$-particle) irreducible diagrams can be performed 
as already shown in \BIPZ\ through appropriate renormalizations. 
It does not involve the introduction of new fundamental objects
and we therefore believe that the corresponding enumeration can be understood
also in terms of S- and R- trees.  

Finally, we have many other solvable matrix models at hand and
we may expect that, for some of them, the underlying combinatorics
can be understood directly in terms of appropriate conjugated
trees. This is already the case for particular two-matrix models,
describing the enumeration of bipartite regular graphs, recently
reformulated in terms of conjugated trees [\xref\BMS-\xref\BFG]. 
The hope is that it may also apply for instance to the case of interacting
two matrix models such as that describing the two-dimensional Ising model 
\ISING\ or hard particle model \HARD\ on random lattices. 
This also suggests to look for a more general picture relating the tree
structure to the (Toda-like) integrable structure of matrix models.

\appendix{A}{Planar maps from the one matrix model}

In this appendix, we recall the derivation of the planar
limit of the integral \partf. For the sake of technical simplicity,
we must take for the potential $V(x)$ a truncated polynomial form
namely
\eqn\useV{ V(x)=\sum_{k=1}^I g_k {x^k \over k} }
The result is then trivially extended so as to involve
arbitrarily many $g_k$'s.  

The planar limit of the integral \partf\ is evaluated in a succession
of standard steps as follows (see e.g. Refs. \DGZ\ and \EY\ for details). 
The first step consists in reducing
the integral \partf\ to one over the eigenvalues of $M$. This is 
readily done by changing variables $M\to (m,U)$, where 
$M=UmU^\dagger$, and $m$ is a real diagonal matrix, while
$U\in U(N)/U(1)^N$ is a unitary matrix
defined up to the right multiplication
by an arbitrary diagonal matrix with entries of the
form $e^{i\phi_j}$ with real phases $\phi_j$, $j=1,2,...,N$. 
The Jacobian of this change 
of variables reads $J=\Delta(m)^2$, where 
$\Delta(m)=\det(m_i^{j-1})\vert_{1\leq i,j\leq N}$ is the Vandermonde 
determinant of $m$. As the integrand of \partf\ only depends on $m$,
the angular variables $U$ may be integrated out, with the result
\eqn\resout{ Z_N(t,\{g_k\})=\int dm \Delta(m)^2 e^{-N{\rm Tr}
\big( {m^2\over 2t}-V(m)\big)} }
up to an overall unimportant multiplicative constant (the volume of the
angular variables' locus).

The second step consists in evaluating the large $N$ limit of 
\resout\ by putting this integral into a more familiar form,
and evaluating it via a saddle-point method.
Indeed, introducing the functional 
\eqn\functS{ S(m)={1\over N}\sum_{i=1}^N \left({m_i^2\over 2t}-V(m_i)\right)
-{1\over N^2}
\sum_{1\leq i\neq j\leq N} {\rm Log}\, |m_i-m_j| }
the integral \resout\ takes the form $Z_N=\int dm e^{-N^2 S(m)}$.
In the limit of large $N$, this integral is dominated by the solution
$m$ to the saddle-point equations $\partial S/\partial m_i=0$, namely
\eqn\sapoS{ {m_i\over t}-V'(m_i)={1\over N} \sum_{j=1\atop j\neq i}^N
{1\over m_i-m_j} }
In particular, we get the planar (genus zero) free energy of the model
as the limit 
\eqn\gezerf{ f(t,\{g_k\})\equiv \lim_{N\to \infty} {1\over N^2}
{\rm Log}\, Z_N(t,\{g_k\})=-S(m)}
evaluated at the solution $m$ to \sapoS.
To further compute this solution, let us introduce the resolvent $\omega_N(\z)$
\eqn\resol{ \omega_N(\z)={1\over N}\sum_{i=1}^N {1\over \z-m_i} }
evaluated on the solution $m$ to \sapoS. 
This function is related to the eigenvalue density $\rho_N(\z)$ defined
as
\eqn\eigdens{ \rho_N(\z)={1\over N}\sum_{i=1}^N \delta(\z-m_i) }
also evaluated on the solution $m$ to \sapoS. More precisely
we have $\rho_N(\z)=(\omega(\z+i0)-\omega(\z-i0))/(2i\pi)$, so
the density of eigenvalues is concentrated on the (real) singularities of $\omega_N$. 
By the definition \resol, it is easy to see that $\omega_N(\z)\sim 1/\z$ when $\z$ is large,
expressing the normalization of the density $\int_{\IR} dx \rho_N(x)=1$.
Let us now 
multiply both sides of \sapoS\
by $1/(\z-m_i)$ and sum over $i=1,2,...,N$. We find the following simple 
quadratic differential equation for $\omega_N(\z)$
\eqn\quadiff{ {1\over N} {d\omega_N(\z)\over d\z}=\omega_N(\z)^2 -\left({\z\over t}-V'(\z)
\right) \omega_N(\z)+P_N(\z)} 
where $P_N(\z)$ is defined by
\eqn\defpn{P_N(\z)= {1\over t}-{1\over N} \sum_{i=1}^N {V'(m_i)-V'(\z)\over m_i-\z} }
Note that, as $V'$ is a polynomial of degree $I-1$, $P_N$ is itself a polynomial of
degree $I-2$.
In the large $N$ limit, the left hand side of \quadiff\ is negligible, and we may therefore 
obtain the large $N$ resolvent $\omega(\z)\equiv \lim_{N\to \infty}\omega_N(\z)$
as the solution of a quadratic equation, in the form
\eqn\forfin{ \omega(\z)={1\over 2}\left({\z\over t}-V'(\z)\pm 
\sqrt{({\z\over t}-V'(\z))^2-4 P(\z)}\right) }
where $P$ denotes the large $N$ limit of $P_N$ (it is also a polynomial of degree $I-2$), 
and the sign $\pm$ is selected
by requiring that the large $\z$ asymptotics of $\omega(\z)$ read $\omega(\z)\sim 1/\z$.

In a third step, the planar resolvent $\omega(\z)$ is further completely fixed by making the
standard ``one-cut" hypothesis, namely that $\omega(\z)$ only has a square root
singularity of the form $\sqrt{(\z-a)(\z-b)}$ where $a$ and $b$ are real, with say $a<b$. 
This expresses the fact that the planar
eigenvalue density $\rho(\z)\equiv \lim_{N\to \infty} \rho_N(\z)$ is assumed to have 
a compact support made of {\it a single real interval} $[a,b]$. 
Picking say $g_I$ to be positive, the sign in \forfin\ is fixed to be $+$, and the 
resolvent takes the form
\eqn\resoform{ \omega(\z)={1\over 2}\big({\z\over t}- V'(\z)+G(\z)\sqrt{(\z-a)(\z-b)} \big) }
where $G(\z)$ is a polynomial of degree $I-2$. Writing that $\omega(\z)\sim 1/\z$ at large $\z$
now determines all coefficients of $G$ and both $a$ and $b$.
More precisely, we have
\eqn\valG{ G(\z)=\left[ {V'(\z)-{\z\over t}\over \sqrt{(\z-a)(\z-b)}} \right]_+ }
where we must expand the bracket as a power series of $\z$ at $\infty$, and the
subscript $+$ indicates that we must retain only the polynomial part of this Laurent series.
Denoting moreover by $H(\z)=(V'(\z)-{\z\over t})/\sqrt{(\z-a)(\z-b)}$ the Laurent series
in the bracket, we may express directly 
\eqn\diromeg{\omega(\z)=-{1\over 2} [H(\z)]_- \sqrt{(\z-a)(\z-b)}}
where the subscript $-$ indicates that we only retain negative powers of $\z$ in the 
Laurent series.
Using the Cauchy formula around the infinity in the complex plane, we may express
the coefficients of $[H(\z)]_-=\sum_{m\geq 1} \z^{-m} H(\z)\vert_{\z^{-m}}$ as
\eqn\obtcof{ H(\z)\vert_{\z^{-m}}= {1\over 2i\pi} \oint w^{m-1} dw 
{V'(w)-{w\over t}\over \sqrt{(w-a)(w-b)}} }
This integral may be drastically simplified if we change variables to $z$:
\eqn\chgvar{ w=Q(z)\equiv z+S+{R\over z} }
and pick $R$ and $S$ so that the square root disappears.
Indeed, 
\eqn\calclong{ (w-a)(w-b)=z^2 +(2S-a-b)(z+{R\over z})+(S-a)(S-b)+2R+{R^2\over z^2} }
is a perfect square iff 
\eqn\defSR{ \eqalign{
S&={a+b\over 2} \cr
R&= \left({b-a\over 4}\right)^2 \cr}}
in which case $\sqrt{(w-a)(w-b)}=z-R/z$ around the infinity. The change of variables $w\to z$
then leads to
\eqn\leadchavar{ H(\z)\vert_{\z^{-m}}= {1\over 2i\pi} \oint {dz \over z}
\left( V'(Q(z))-{Q(z)\over t}\right) Q(z)^{m-1} }  
For the first few values of $m=1,2,3,4$, this gives
\eqn\firfewval{ \eqalign{
H(\z)\vert_{\z^{-1}}&= V'(Q)\vert_{z^0} -{S\over t} \cr
H(\z)\vert_{\z^{-2}}&= 2\big( V'(Q)\vert_{z^{-1}} -{R\over t} \big)+S H(\z)\vert_{\z^{-1}} \cr
H(\z)\vert_{\z^{-3}}&= 2V'(Q)\vert_{z^{-2}}+2SH(\z)\vert_{\z^{-2}}+(2R-S^2)H(\z)\vert_{\z^{-1}} \cr
H(\z)\vert_{\z^{-4}}&= 2V'(Q)\vert_{z^{-3}}+3SH(\z)\vert_{\z^{-3}}+3(R-S^2)H(\z)\vert_{\z^{-2}}+(S^3-3 RS)
H(\z)\vert_{\z^{-1}} \cr }}
In the second, third and fourth lines of \firfewval, we have used the symmetry $z\leftrightarrow R/z$
of the change of variables to express that $f(Q)\vert_{z^{-m}}=R^m f(Q)\vert_{z^m}$
for any Laurent series $f$, and all $m\geq 0$, in order to reexpress the result in terms
mostly of $V'(Q)$. 
We now simply have to substitute the above expressions for the coefficients
of $H$ into the formula \diromeg\ for $\omega(\z)$, and impose that $\omega(\z)=1/\z+O(1/\z^2)$ at
infinity. This gives respectively for the order $\z^0$ and $\z^{-1}$ terms the two equations
\eqn\twoeq{\eqalign{
{S\over t}- V'(Q)\vert_{z^0}&= 0\cr
{R\over t}-V'(Q)\vert_{z^{-1}} &= 1 \cr}}
which are nothing but \SR.

The planar resolvent $\omega(\z)$ may also be used as the generating function for the
quantities $\theta_k=\lim_{N\to \infty} {1\over N}\langle {\rm Tr}(M^k) \rangle$
that enumerate the (possibly disconnected) planar maps with one special $k$-valent vertex
and arbitrary regular $m$-valent ones weighted by $g_m$, $m=1,2,3,...$ and with a weight
$t$ per edge. Indeed, the definition of $\omega(\z)$ entails that
\eqn\genom{ \omega(\z)=\sum_{k=0}^\infty \z^{-k-1} \lim_{N\to \infty}{1\over N}
{\rm Tr}(m^k)=\sum_{k=0}^\infty \z^{-k-1} \theta_k}
where $m$ stands for the solution to the saddle-point equations \sapoS.  
In particular, let us use the direct expression \diromeg\ to compute the coefficients
$\theta_1$ and $\theta_2$ corresponding to one and two-leg diagrams respectively. 
Substituting \firfewval\ into \diromeg, we get
\eqn\twothets{\eqalign{
\theta_1 &= S -V'(Q)\vert_{z^{-2}} \cr
\theta_2 &= R+S^2-V'(Q)\vert_{z^{-3}}-2 S V'(Q)\vert_{z^{-2}} \cr}}
Finally, we note that the one-leg diagrams counted by $\theta_1$ are always connected
hence $\theta_1=\Gamma_1$, while the two-leg ones may be disconnected, and we must
write $\Gamma_2=\theta_2-\theta_1^2$. This finally leads to Eqs. \eqGamma. 

A first remark is in order. 
The quantity $R$ may be directly expressed as a connected
matrix average, namely
\eqn\connmat{ R= \theta_{1,1} -\theta_1^2 }
where 
\eqn\defoneone{ \theta_{1,1}=\lim_{N\to \infty} {1\over N^2} 
\langle {\rm Tr}(M)\,  {\rm Tr}(M)\rangle}
where the average is computed by attaching two external legs to connected planar maps
and those two legs {\it do not have to lie in the same face} (as they do in $\Gamma_2$).
To prove this, the simplest way is to use the linear term in the potential
to generate insertions of Tr$(M)$ by differentiating \wrt $g_1$. 
For instance,
the connected average $\Gamma_{1,1}\equiv \theta_{1,1} -\theta_1^2$ is easily obtained by
differentiating $\theta_1$ \wrt $g_1$:
\eqn\caltetone{\eqalign{ \Gamma_{1,1}&=
{\partial \theta_1 \over \partial g_1}\cr
&= {\partial S\over \partial g_1} 
-{\partial V'(Q)\vert_{z^{-2}}\over \partial g_1}\cr
&=(1-V''(Q)\vert_{z^{-2}}){\partial S\over \partial g_1}- V''(Q)\vert_{z^{-1}} 
{\partial R\over \partial g_1}\cr}}
Let us now use the second equation of \twoeq\ to express the $1$ in the first factor
as $1=R/t-V'(Q)\vert_{z^{-1}}$, and upon
differentiating the first line of \twoeq\ \wrt $g_1$ and multiplying by $R$, 
we may eliminate $(R/t) \partial S/\partial g_1-V''(Q)\vert_{z^{-1}}
\partial R/\partial g_1=R+RV''(Q)\vert_{z^{0}} \partial S/\partial g_1$, to get 
\eqn\gamnul{ \Gamma_{1,1}=R +(RV''(Q)\vert_{z^0}-V'(Q)\vert_{z^{-1}}-V''(Q)\vert_{z^{-2}})
{\partial S\over \partial g_1} }
Finally, it is easy to see that the prefactor of $\partial S/\partial g_1$ vanishes,
as it is nothing but the residue of a total derivative: 
$-{d\over dz} \big(zV'(Q)\big)\vert_{z^{-1}}=0$.
We finally get the desired result $\Gamma_{1,1}=R$.

A last remark is in order. When all $g_{2k-1}=0$, $k=1,2,3,...$,  
the partition function \partf\ counts possibly disconnected Eulerian maps,
namely with only regular vertices of even valence. In our present formulation,
we see that the complete potential ${x^2\over 2t} -V(x)$ is an even function of $x$,
which simplifies the results considerably. Indeed, we may immediately infer that
the planar density of eigenvalues is also even, so that $a+b=0$, henceforth $S=0$.

\appendix{B}{Proof of the relation $R_1+2 R_0=(V'(Q)|_{z^{-2}})^2$}

\fig{A schematic representation of an S$_k$-tree (on the left)
and of an ${\bar{\rm S}}_l$-tree (on the right). We represent only
the buds and leaves which play a role in the evaluation of
the depth of the marked (by a cross) leaf or bud. We also
indicate the closing arches in dashed lines. 
 }{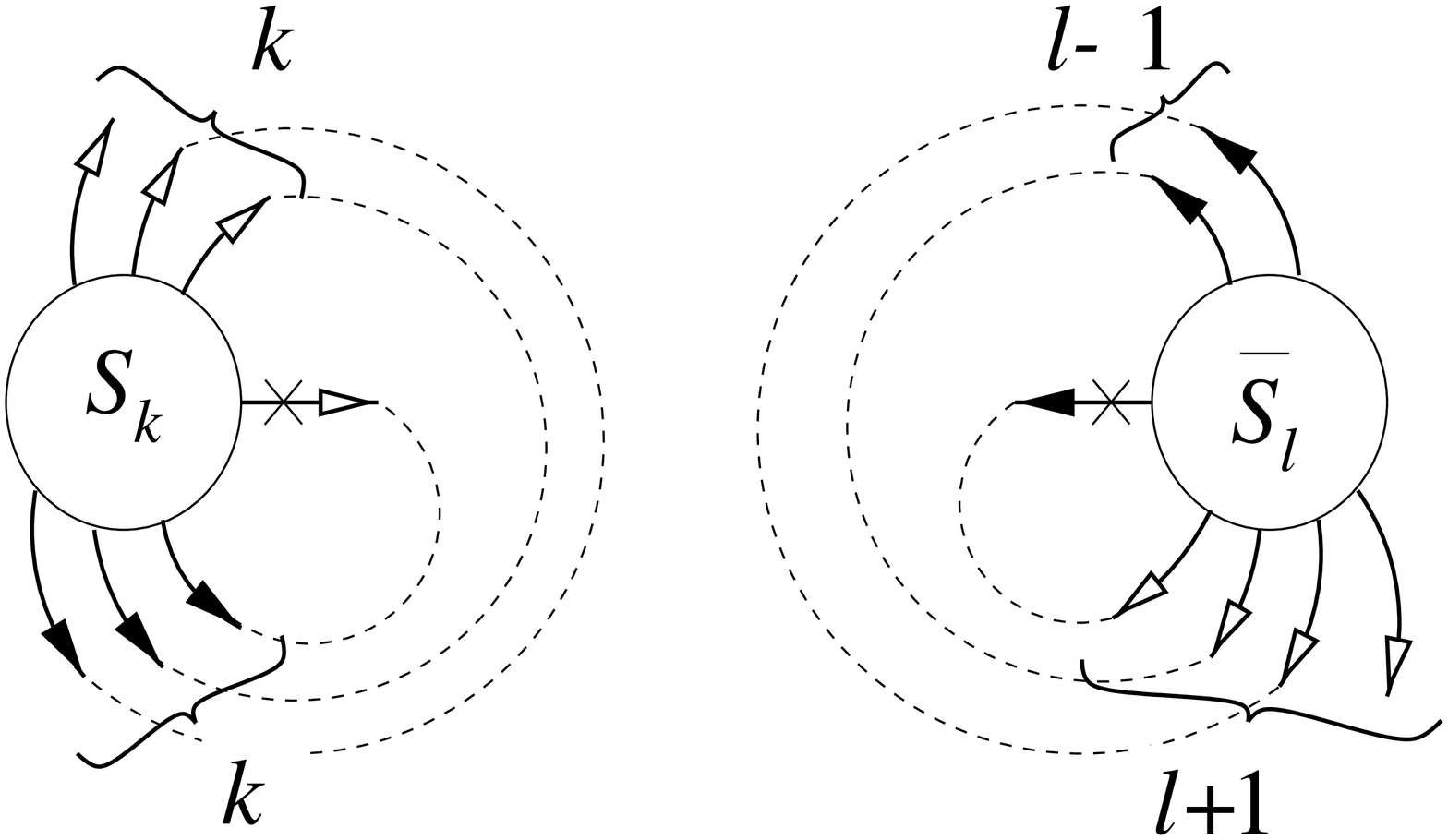}{9.cm}
\figlabel\SkSl

Let us introduce refined definitions of S-trees as follows.
We define an S$_k$- (resp. an ${\bar{\rm S}}_l$- ) tree as an
unrooted S-tree with a marked leaf (resp. a marked bud) of depth $k$
(resp. $l$). An illustration of these definitions is given in
Fig. \SkSl. 
We denote by $S_k$ (resp. ${\bar S}_l$) the corresponding generating 
functions. As the number of leaves and buds of given depth $k\geq 1$
are the same in any unrooted S-tree, we have $S_k={\bar S}_k$ for all
$k\geq 1$. For $k=0$ we have ${\bar S}_0=0$ while $S_0=\Gamma_1$
as there is a unique depth $0$ leaf in any unrooted S-tree.
Moreover we have $S=\sum_{k\geq 0} S_k$ as the choice of root for
an S-tree is that of a leaf of arbitrary depth. Using the first line
of \eqGamma\ (for which we have already given a combinatorial proof
in Section 3.3), we may express 
\eqn\vprime{V'(Q)|_{z^{-2}}=S-\Gamma_1=\sum_{k\geq 1} S_k=\sum_{k\geq 1}
{\bar S}_k}
and therefore
\eqn\vcarre{(V'(Q)|_{z^{-2}})^2=\sum_{k\geq 1} S_k{\bar S}_k +2 \sum_{k>l \geq 1}
S_k{\bar S}_l}
Let us now identify the right hand side of this equation with $R_1+2 R_0$.

The quantity $S_k{\bar S}_l$ counts all trees obtained by gluing into an edge
the marked leaf of an S$_k$-tree with the marked bud of 
an ${\bar{\rm S}}_l$-tree. The resulting tree is easily seen to be an unrooted
R-tree (of total charge $2$) with a particular marked edge. This
edge separates the unrooted R-tree into a rooted S-subtree (the former
S$_k$-tree with its marked leaf replaced by a root) and a part
of charge $2$ (obtained by removing the marked bud of the ${\bar{\rm S}}_l$-tree).
Let us show that the S-subtree is maximal hence the marked edge connects
it to the core of the unrooted R-tree. Indeed, the maximal S-tree containing
this S-subtree is connected to the core through a $0$-$2$ edge. This edge
separates the ${\bar{\rm S}}_l$-tree into a piece of charge $2$ and one
of charge $-1$, which in an unrooted S-tree is possible only if the
part of charge $-1$ is reduced to a bud. This identifies the above $0$-$2$ edge
with the marked one, and proves that the S-subtree is maximal. Note moreover that the core
of the R-tree is actually obtained by gluing the cores of all the R-subtrees
of the ${\bar{\rm S}}_l$-tree originating from the vertex to which the marked bud
is attached.
The quantity $S_k{\bar S}_l$ counts unrooted R-trees with a singularized
maximal S-subtree and particular depth restrictions. 

\fig{A schematic view of the unrooted R-tree resulting from the
connection of an S$_k$-tree with an ${\bar{\rm S}}_l$-tree via
their marked leaf and bud. The positions of the two unmatched leaves
in the closing procedure depend on whether (a) $k=l$ (one on each side)
(b) $k>l$ (both on the left) or (c) $k<l$ (both on the right).}{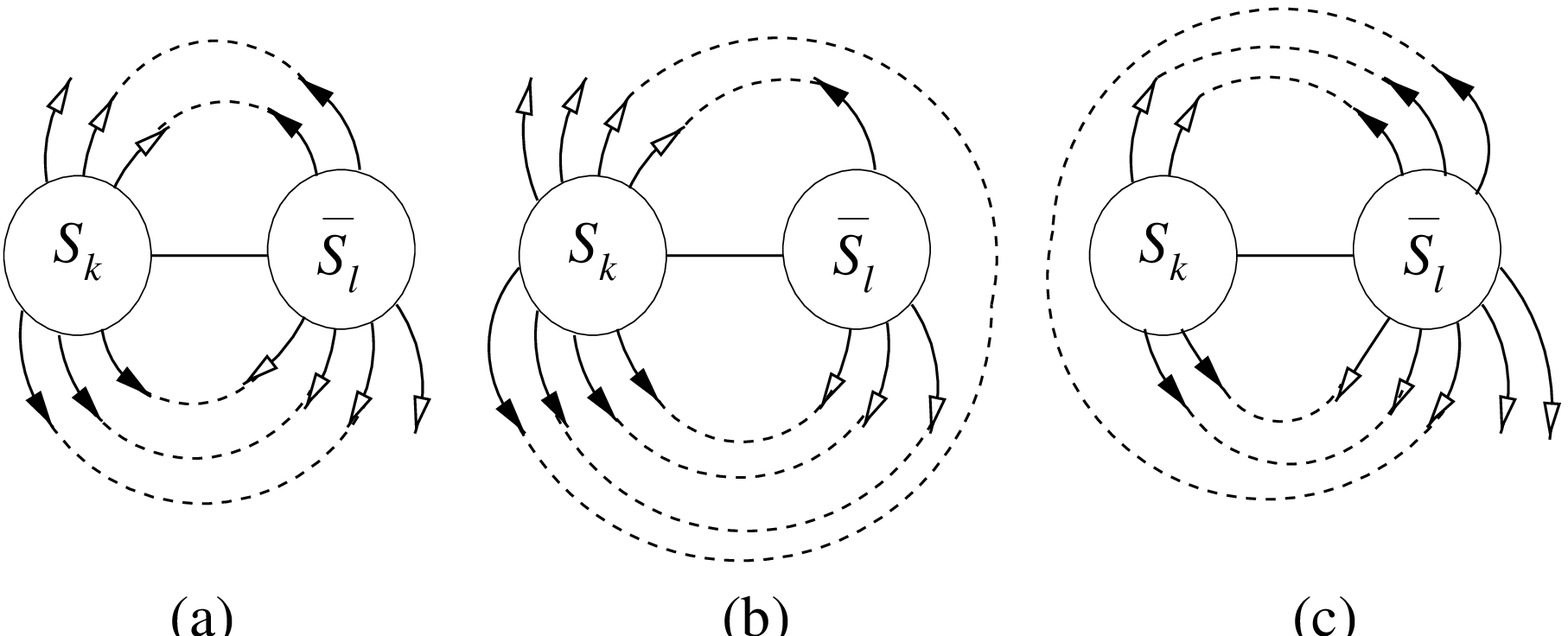}{12.cm}
\figlabel\SkSlmatch

Now it is easy to check that if $k=l$, the bud-leaf matching procedure 
leaves us with {\it one unmatched leaf on each side} 
(see Fig. \SkSlmatch). On the other hand, for $k\neq l$, the two unmatched   
leaves are on the same side, which is that of the singularized 
maximal S-subtree iff $k>l$.  
The right hand side of \vcarre\ therefore counts the number of
unrooted R-trees with a singularized maximal S-subtree and with a
marked depth $0$ leaf on it. This also counts the unrooted R-trees
twice if their two depth $0$ leaves lie outside the core (i.e.
in either the same or two distinct maximal S-subtrees), and once
if exactly one depth $0$ leaf lies outside the core (i.e. is
in a maximal S-subtree). This is nothing but $2R_0+R_1$. QED.

\listrefs
\end